\documentclass[epj]{svjour}
\usepackage{latexsym}
\usepackage{graphicx}
\usepackage{dcolumn}
\usepackage{bm}

\usepackage[utf8]{inputenc}
\usepackage[T1]{fontenc}
\usepackage{mathptmx}
\usepackage{etoolbox}

\usepackage{amssymb}
\usepackage{amsmath}
\usepackage{mathtools}

\usepackage{mdframed}

\usepackage{xcolor}

\newcommand{\homega}{\bar \omega}

\newcommand{\mr}{\mathrm}

\newcommand{\bra}[1]{\left\langle #1 \right\vert}
\newcommand{\ket}[1]{\left\vert #1 \right\rangle}

\makeatletter
\def\@email#1#2{%
 \endgroup
 \patchcmd{\titleblock@produce}
  {\frontmatter@RRAPformat}
  {\frontmatter@RRAPformat{\produce@RRAP{*#1\href{mailto:#2}{#2}}}\frontmatter@RRAPformat}
  {}{}
}%
\makeatother
\usepackage{graphics}
\graphicspath{{.}{./.}}
\begin{document}
\title{Quantum spin-flavour memory of ultrahigh-energy neutrino}

\author{P. Kurashvili\inst{1} 
\and L. Chotorlishvili\inst{2}
\and K. A. Kouzakov\inst{3} 
\and A. I. Studenikin\inst{4, 5}
}                     
%
%
\institute{ National Centre for Nuclear Research, Warsaw 00-681, Poland
\and  
Department of Physics and Medical Engineering, Rzesz\'ow University of Technology, 35-959 Rzesz\'ow, Poland
\and 
Department of Nuclear Physics and Quantum Theory of Collisions, 
Faculty of Physics, Lomonosov Moscow State University, Moscow 119991, Russia
\and
Department of Theoretical Physics, Faculty of Physics,
Lomonosov Moscow State University, Moscow 119991, Russia
\and
Joint Institute for Nuclear Research, Dubna 
141980, Moscow Region, Russia
}
\date{Received: date / Revised version: date}
%
\abstract{
There are two types of uncertainties related to the measurements done 
on a quantum system: statistical and those related to non-commuting 
observables and incompatible measurements. 
The latter indicates the quantum system's inherent nature and is in the scope 
of the present study. 
We explore uncertainties related to the interstellar 
ultrahigh-energy neutrino and introduce a novel concept: 
quantum spin-flavour memory.  
Advanced uncertainty measures are entropic measures, and the effect of the 
quantum memory reduces the uncertainty.  
The problem in question corresponds to a real physical event:   high-energy 
Dirac neutrinos emitted by some distant source and propagating towards 
the earth. 
The neutrino has a finite magnetic moment and interacts with both deterministic and stochastic interstellar magnetic fields. 
To describe the effect of a noisy environment, we exploit 
the Lindblad master equation for the neutrino density matrix. 
Quantum spin-flavour memory we quantify in terms of the generalized Kraus's 
trade-off relation. 
This trade-off relation converts to the equality when quantum memory is absent.  
We discovered that while most measures of quantum correlations show their 
irrelevance, the quantum spin-flavour discord is the quantifier of the quantum 
spin-flavour memory. 
\PACS{
      {PACS-key}{describing text of that key}   \and
      {PACS-key}{describing text of that key}
     } 
} 
\maketitle

\section{\label{sec:introduction}Introduction}

In non-relativistic quantum theory, Heisenberg's uncertainty principle 
asserts a  limit to the precision of measuring momentum and coordinate 
of a particle. 
In a broader sense, Heisenberg's uncertainty principle concerns 
any two incompatible measurements of the expectation values of non-commuting 
operators. We point out the recent interest to the relativistic uncertainty 
principle (see 
\cite{todorinov2019relativistic,bishop2019}, and references therein).
The progress in quantum metrology is related to the entropic uncertainty 
relation and entropic measures.  
As distinct from the standard deviations, entropic uncertainty relations 
are independent of the system's states to be measured 
and therefore support the more general formulation of the problem. 
The critical issue for the quantum measurement and entropic measures are quantum 
correlations and entanglement.  
The latter has a certain interplay with thermodynamics. 
For example, entanglement can be exploited as a resource for producing 
quantum work \cite{chotorlishvili2011thermal}. 
Due to the quantum correlations,  measurements performed through the 
classical device on the quantum systems lead to the entropy production 
and the key issue is the spectral property of the 
system \cite{chotorlishvili2010quantum,ugulava2005irreversible}. 
One can recall other additional facts. 
Our interest in the present work is about the uncertainty 
principle in the presence of quantum memory \cite{berta2010uncertainty,tomamichel2011uncertainty,PhysRevB.100.174413,PhysRevA.86.042105,PhysRevA.86.012113,PhysRevA.87.022314,PhysRevA.93.062123,PhysRevA.100.012131,PhysRevLett.110.240402,PhysRevLett.112.050401,PhysRevLett.122.200601,PhysRevX.9.031045}. 
As a physical system under study we consider the ultrahigh-energy neutrino 
in an interstellar magnetic field~\cite{kurashvili2017spin}. The reason is that neutrinos obey dichotomic left-right helicity and trichotomic lepton flavour
(electron, muon, and tau) and when propagating, they can change their type or oscillate. In contrast to the stellar environments, in interstellar space the matter density is very low and neutrinos can be mainly affected by the neutrino magnetic moment interaction with an interstellar magnetic field. As a result, neutrino spin-flavour oscillations can arise in this case, namely the neutrino can change both its flavour and helicity (see Ref.~\cite{kurashvili2017spin} and references therein). Below we formulate a concept of quantum spin-flavour memory for investigating this phenomenon.

We consider a model that can mimic a real physical situation when a flux of
ultrarelativistic flavour neutrinos emitted deep inside a compact massive
astrophysical object (i.e., a core-collapse supernova or a protoneutron object developed from a neutron star merger) propagates to an observer in the Earth.
Suppose that along their path from the source to the Earth neutrinos, that propagate through highly rarefied space with a stochastic magnetic field, also encounter a dense magnetized and rotating astrophysical object (that can be a pulsar, for instance).  The influence of the described above external environments on neutrinos can be considered as a chain of causes of neutrinos' states change. 
Two of these state disturbances can be considered as two independent 
measurements. In particular, in the initial phase of the neutrino  propagation deep inside the dense astrophysical object the neutrino undergoes interaction with dense matter. As a result,  the left-handed neutrinos can be deflected that creates  spin asymmetry in the emitted neutrino flux. This can serve as a "cosmic neutrino filter" inside the astrophysical sources (see \cite{PhysRevD.103.036011} for detailed discussions).
Predominantly, only nearly sterile right-handed neutrinos survive. 
This process can be considered as an effective measurement of neutrino spin done with the spin-one-half operator $\sigma_z$. 
The second "measurement" is provided by the neutrino interaction 
due to its magnetic moment with the external magnetic field of the second 
compact astrophysical object and/or by weak interaction of neutrinos with the 
transversal matter currents which influence neutrinos in the case of possible 
misalignment 
of the direction of neutrinos propagation and the rotation axes of the 
astrophysical object. 
Both these interactions can initiate the neutrino spin oscillations that 
can be also (see \cite{studenikin2004neutrinos} and \cite{PhysRevD.98.113009}) 
increased to the maximum by the corresponding resonances. 
These resonant neutrino spin conversions can be considered as another effective measurement of neutrino spin done with the spin-one-half operator that should not be the same as in the case of the first measurement.  Thus, in general case this second measurement is done by the spin operator  $\sigma_x$.

While "measurements" done on the neutrino and filtering out a particular type of neutrinos occur due to the interaction of neutrino flux with the specific cosmic objects, we exploit projector operators to formulate a problem mathematically. A positive operator-valued measure (POVM) is a set of finite number positive semi-definite operators acting in the Hilbert space. Projecting of neutrino states on the particular states mimics the filtering procedure described above. In its most straightforward form, the procedure can be illustrated for qubits. Let $\vert\varphi\rangle=a\vert 0\rangle+b\vert 1\rangle$ be the state of a single qubit. The action of the projector $\hat\Pi=\vert n\rangle\langle n\vert$, (in what follows due to the physical interest $\vert n\rangle$ are eigenstates either of $z$ or $x$ components of the neutrino spin), retains only desired filtered post-measurement state $\vert\psi\rangle=\left[ \sqrt{\langle\varphi\vert\hat\Pi\vert\varphi\rangle}\right] ^{-1/2}\hat\Pi\vert\varphi\rangle$.  In the same sense, POVMs can be applied to the neutrino states to describe filtering out a particular type of neutrinos. The elimination of a particular type of neutrinos occurs due to the interaction with specified cosmic objects.

Let $H(\sigma^z)=-\sum\limits \sigma^z_i\log_2 \sigma^z_i$ and 
$H(\sigma^x)=-\sum\limits \sigma^x_i\log_2 \sigma^x_i$
be Shannon's entropies for the outcomes of two incompatible measurements 
done on the quantum spin-1/2 observables $\hat{\sigma}^z$ and 
$\hat{\sigma}^x$.  
Here 
$\sigma^z_i=\vert\langle \sigma^z_i\vert\Psi\rangle\vert^2$ 
and $\sigma^x_i=\vert\langle \sigma^x_i\vert\Psi\rangle\vert^2$ 
are the probability distributions of the measurement outcomes
(see the seminal work of David Deutsch \cite{PhysRevLett.50.631}), where the state of the system is given by $\vert\Psi\rangle$ and 
$\vert\sigma_i^z\rangle$, $\vert\sigma_i^x\rangle$
are the eigenstates of the observables. 
The Kraus's trade-off relation proved by Maassen and Uffink reads    
\cite{kraus1987complementary,maassen1988generalized}: 
\begin{eqnarray}\label{trade-off relation1}
H(\sigma^z)+H(\sigma^x)\geqslant -2\log_2c(\sigma^z,\sigma^x),
\end{eqnarray}
where $c(\sigma^z,\sigma^x)\equiv \max_{ij} 
\vert\langle \sigma^z_i \vert \sigma^x_j\rangle\vert$, 
and $\vert\sigma^z_i\rangle, 
\vert\sigma^x_j\rangle$ are the eigenvectors of $\hat{\sigma}^z$ and 
$\hat{\sigma}^x$, respectively. 
We focus on the bipartite system $\hat\rho_{\sigma\nu}$. 
In what follows, $\nu$ defines the neutrino flavour subspace in the 
approximation of two neutrino generations ($\nu=\nu_e,\nu_\mu$) and $\sigma$
defines the spin subspace. 
In the work \cite{PhysRevA.86.042105} the theorem was proved 
that quantifies the correction to the Kraus's trade-off relation
\cite{kraus1987complementary,maassen1988generalized}. 
We adopt this theorem to the ultrahigh-energy neutrino problem: 
\begin{eqnarray}\label{correction to the The Kraus}
 S(R\vert\nu)+S(Q\vert\nu)\geqslant -2\log_2 c(R, Q)+S(\sigma\vert\nu)+
 \max\left\lbrace 0, D_{\sigma}(\hat\rho_{\sigma\nu})-J_{\sigma}(\hat\rho_{\sigma\nu})\right\rbrace. 
\end{eqnarray}
Here $S(R\vert\nu)$ and $S(Q\vert\nu)$ are the conditional von Neumann entropies of the post-measurement states 
\begin{eqnarray}\label{correction to the The Kraus2}
S(R\vert\nu)=S(\hat\rho_{R\nu})-S(\hat\rho_{\nu}),\qquad 
S(Q\vert\nu)=S(\hat\rho_{Q\nu})-S(\hat\rho_{\nu}),
\end{eqnarray}
i.e., $S(\hat\rho_{R\nu})=-\hat\rho_{R\nu}\ln(\hat\rho_{R\nu})$, where $\hat\rho_{R\nu}$ is the post-measurement density matrix, and two measurements are done on the components of spin $R\equiv \sigma_z$, meaning that POVMs are constructed from eigenvectors $\vert\sigma^z_j\rangle\equiv\vert r\rangle$:
\begin{eqnarray}\label{added equation one}
 \hat\rho_{R\nu}=\sum\limits_r\vert r \rangle\langle r \vert_R\otimes Tr_\sigma\left\lbrace (\Gamma(r)\otimes I_{\nu})\hat\rho_{\sigma\nu}\right\rbrace. 
\end{eqnarray}
The same is valid for $\hat\rho_{Q\nu}$. 
Post-measurement density matrix $\hat\rho_{Q\nu}$ is obtained after 
measuring $x$ component of the neutrino spin and can be constructed similarly 
to Eq. (\ref{added equation one}) by considering $Q\equiv \sigma_x$.
The classical correlation term in 
Eq. (\ref{correction to the The Kraus}) is defined as follows:
\begin{eqnarray}\label{The classical correlation term}
 J_{\sigma}(\hat\rho_{\sigma\nu})=\max_{\Gamma (x)}I(X,\nu),
\end{eqnarray}
where the mutual information 
\begin{eqnarray}\label{the mutual information}
 I(X,\nu)=S(X)+S(\nu)-S(X\nu)
\end{eqnarray}
is calculated for the post-measurement state  
\begin{eqnarray}\label{post-measurement state}
 \hat\rho_{X\nu}=\sum\limits_x\vert x \rangle\langle x \vert_X\otimes Tr_\sigma\left\lbrace (\Gamma(x)\otimes I_{\nu})\hat\rho_{\sigma\nu}\right\rbrace. 
\end{eqnarray}
Here $\Gamma(x)$ is the set of all positive operator-valued measures (POVMs) 
acting on the spin subspace. Throughout the whole work all
the measurements mentioned in the text concern the spin subspace. The last term 
in Eq. (\ref{correction to the The Kraus}) quantifies the quantum discord:
\begin{eqnarray}\label{quantum discord}
&&  D_{\sigma}(\hat\rho_{\sigma\nu})=I(\sigma,\nu)-J_{\sigma}(\hat\rho_{\sigma\nu}).
\end{eqnarray}

We note that the right-hand side of 
Eq. (\ref{correction to the The Kraus})  quantifies the lower bound of 
the uncertainty and contains three terms.
The first term quantifies the incompatibility of POVMs 
acting on the neutrino's spin. 
The second term $S(\sigma\vert\nu)$ is the conditional entropy. 
The third term $\max\left\lbrace 0, D_{\sigma}(\hat\rho_{\sigma\nu})-J_{\sigma}(\hat\rho_{\sigma\nu})\right\rbrace$ is related to the quantum discord. 
The sum of all these three terms on the right-hand side of
Eq. (\ref{correction to the The Kraus}) defines the lower bound of the uncertainty. We are interested in exploring the case when inequality holds and the lower bound of uncertainty is reduced.

The spin of the neutrino possessed by Alice 
(the cosmic $\sigma_z$ and $\sigma_x$ measurements)
and the flavour that belongs to Bob (the observer at the Earth)
are not entirely independent quantum numbers. 
Bob inherits spin-flavour quantum memory that reduces uncertainties of Alice's 
two measurements done on the spin subsystem. 
We consider the case when after the first measurement, the system 
undergoes dissipative evolution through the Lindbladian channel. 
We note that dissipative evolution is relevant for the neutrino evolution
in the interstellar space. 
For details and physical circumstances, we refer to two recent works
\cite{PhysRevD.103.036011,kurashvili2021coherence}. 
Here we would like to explore robustness of the quantum spin-flavour memory 
with respect to the Lindbladian evolution.  
It should be noted that there are quantum information studies of neutrino 
oscillations (see, for instance, \cite{banerjee2015quantum}), 
but in contrast to our present work they typically do not involve 
neutrino spin and spin-flavour transitions and dissipative effects.

The work is organized as follows: in 
Section \ref{sec:model} we specify the model, in Section \ref{sec:solution}
we discuss the solution of the Lindblad equation, in Section 
\ref{sec:measures} we
explore the entropic measures, in Section \ref{sec:results} we analyse results and 
conclude the work.


\section{\label{sec:model}Model}
We define the measurement register and the  memory through the relations:
\begin{eqnarray}\label{Quantum memory1}
&&\hat{\rho}_{R\nu}=
\sum\limits_{n}|\psi_{n}\rangle\langle \psi_{n}|\otimes I_{\nu} \hat{\rho}_{\sigma\nu}
|\psi_{n}\rangle\langle \psi_{n}|\otimes I_{\nu},
\nonumber\\
&&\hat{\rho}_{Q\nu}=
\sum\limits_{n}|\phi_{n}\rangle\langle \phi_{n}|\otimes I_{\nu} \mathcal{\hat{L}}(\hat{\rho}_{R\nu})|\phi_{n}\rangle\langle \phi_{n}|\otimes I_{\nu}.
\end{eqnarray}
The identity operator $I_{\nu}$ acts on the flavour subspace $\nu$, 
and $|\psi_{1}\rangle=|1\rangle,~~|\psi_{2}\rangle=|0\rangle$, 
$|\phi_{1,2}\rangle=\frac{1}{\sqrt{2}}(|0\rangle \pm|1\rangle)$,  
are the eigenfunctions of the neutrino spin operators 
$\hat \sigma_{z}$, $\hat \sigma_{x}$, 
$c\left(\hat\sigma_x,\hat\sigma_y\right)
=\max\Vert \sqrt{\Gamma_x}\sqrt{\Gamma_y}\Vert $ 
quantifies incompatibility of POVMs acting on the neutrino's spin, 
and $\hat{\varrho}_{\sigma\nu}=\mathcal{\hat{L}}(\hat{\rho}_{\sigma\nu})$ is 
the Lindbladian trace preserving evolution tackled in the Novikov's form 
\cite{novikov1965functionals}. 
Despite abstract character, projection operations have clear physical meaning: 
the neutrino spins are aligned along (transversely to) the neutrino propagation 
direction after the first (second) measurement. 
This polarization effect can be described through POVM spin projectors 
\cite{kaspi2017magnetars}. 
Neutrino flux reaching us from such distant sources is affected by galactic and 
extragalactic magnetic fields. 
These weak fields can exert a dissipative effect \cite{stebbins2019new}. 
In what follows, we describe this effect through the dissipative 
trace-preserving Lindblad equation 
(for more details we refer to \cite{PhysRevD.103.036011}).

We define two helicity basis states for the Dirac neutrino  
\cite{song2018quantifying,PhysRevLett.117.050402,kurashvili2017spin} 
$\vert\nu_{1,s=\pm 1}\rangle$, $\vert\nu_{2,s=\pm 1}\rangle$ with masses $m_1$ and $m_2$. Using the helicity basis states 
we define the flavour basis 
\begin{eqnarray}\label{flavour basis}
&&\vert\nu_{e}^{R,L}\rangle=\vert\nu_{1,s=\pm 1}\rangle\cos\theta_\nu
+\vert\nu_{2,s=\pm 1}\rangle\sin\theta_\nu,\nonumber\\
&&\vert\nu_{\mu}^{R,L}\rangle=\vert\nu_{1,s=\pm 1}\rangle\sin\theta_\nu
+\vert\nu_{2,s=\pm 1}\rangle\cos\theta_\nu.
\end{eqnarray}
Here $\theta_\nu$ is the neutrino mixing angle.
The Hamiltonian of the system comprises the several terms:
\begin{eqnarray}\label{Hamiltonian of the problem}
&& \hat{H}_{eff}=\hat{H}_{vac}+\hat{H}_{mat}+\hat{H}_{B}.
\end{eqnarray}
Here $\hat{H}_{vac}$ is the vacuum part and terms 
$\hat{H}_{vac}$, $\hat{H}_{B}$ describe neutrino interaction with the matter 
and magnetic field, respectively 
(see for more details \cite{kurashvili2017spin}).
The vacuum part $\hat{H}_{vac}$ explicitly has the form %
\begin{equation}
\label{eq:HamiltonianFlavorRepresentation}
{\hat H}_{vac}=\omega
\begin{pmatrix}
-\cos 2 \theta_{\nu} & 0 & \sin 2 \theta_{\nu} & 0
\\
0 & -\cos 2 \theta_{\nu} & 0 & \sin 2 \theta_{\nu}
\\
\sin 2 \theta_{\nu} & 0 & \cos 2\theta_{\nu} & 0
\\
0 & \sin 2 \theta_{\nu} & 0 & \cos 2\theta_{\nu}
\end{pmatrix},
\end{equation}
with
\begin{equation}
\label{eq:DeltaM}
\omega= \frac{\Delta m^2}{4 E_\nu}, \qquad \Delta m^2 =m_2^2 - m_1^2,
\end{equation}
and $E_\nu$ being the neutrino energy. The neutrino-matter interaction is described by the Hamiltonian
\begin{equation}
\label{eq:InteractionHamiltonian}
\hat{H}_{mat}=
 \frac {G_F }{\sqrt{2}}
\begin{pmatrix}
0 & 0 & 0 & 0
\\
0 & n^{(\nu_{e})}_{eff} & 0 & 0
\\
0 & 0 & 0 & 0
\\
0 & 0 & 0 & n^{(\nu_{\mu})}_{eff}
\end{pmatrix},
\end{equation}
where $n^{(\nu_{e})}_{eff}=  n_e  - {n_n}/{2}$ and $n^{(\nu_{\mu})}_{eff}= - {n_n}/{2}$, with the Fermi constant $G_F$, the net electron density
$n_e=n_{e^-}-n_{e^+}$ and the neutron density $n_n$.
The Hamiltonian of the neutrino interaction with a magnetic field in the
flavour basis can be presented as~\cite{fabbricatore2016neutrino}
%
\begin{equation}
\label{eq:H_EM}
\hat{H}_{B}=
\begin{pmatrix}
\displaystyle -\left(\frac{\mu}{\gamma}\right)_{ee} {B_\parallel} &&
\mu_{ee}B_{\perp} &&
\displaystyle -\left(\frac{\mu}{\gamma}\right)_{ e\mu}{B_\parallel} &&
\mu_{e\mu}B_\perp
\\
\mu_{ee}B_\perp &&
\displaystyle -\left(\frac{\mu}{\gamma}\right)_{ee}{B_\parallel} &&
\mu_{e\mu} B_\perp &&
\displaystyle -\left(\frac{\mu}{\gamma}\right)_{e\mu}{B_\parallel}
\\
\displaystyle -\left(\frac{\mu}{\gamma}\right)_{e\mu}{B_\parallel} &&
\mu_{e\mu}B_{\perp} &&
\displaystyle -\left(\frac{\mu}{\gamma}\right)_{\mu \mu}{B_\parallel}
&&
\mu_{\mu \mu} B_\perp
\\
\mu_{e\mu}B_{\perp} &&
\displaystyle -\left(\frac{\mu}{\gamma}\right)_{e\mu}{B_\parallel} &&
\mu_{\mu \mu}B_\perp &&
\displaystyle -\left(\frac{\mu}{\gamma}\right)_{\mu\mu}{B_\parallel}
\end{pmatrix},
\end{equation}
%
where $B_\parallel$ and $B_\perp$ are the parallel and transverse magnetic-field components with respect to the neutrino velocity, and the magnetic moments $\tilde{\mu}_{\ell\ell'}$ and $\mu_{\ell\ell'}$ ($\ell,\ell'=e,\mu$) are related to those in the mass representation $\mu_{jk}$ ($j,k=1,2$) as follows:
\begin{align}
\label{eq:MuPrime}
\mu_{ee}&=\mu_{11} \cos^2 \theta_{\nu} +\mu_{22} \sin^2 \theta_{\nu}
+\mu_{12} \sin 2\theta_{\nu},
\nonumber
\\
\mu_{e\mu}&=\mu_{12}\cos 2\theta_{\nu} + \frac{1}{2}
\left( \mu_{22} - \mu_{11}\right)\sin 2\theta_{\nu},
\\
\mu_{\mu\mu}&=\mu_{11} \sin^2 \theta_{\nu}
+\mu_{22} \cos^2 \theta_\nu-\mu_{12} \sin 2\theta_{\nu},
\nonumber
\end{align}
and
\begin{align}
\label{eq:MuTilde}
-\left(\frac{\mu}{\gamma}\right)_{ee} &=
\frac{\mu_{11}}{\gamma_{1}}\,\cos^2 \theta_{\nu} +
\frac{\mu_{22}}{\gamma_{2}}\,\sin^2 \theta_{\nu} +
\frac{\mu_{12}}{\gamma_{12}}\,\sin 2\theta_{\nu},
\nonumber
\\
-\left(\frac{\mu}{\gamma}\right)_{e\mu} &=
\frac{\mu_{12}}{\gamma_{12}}\,\cos 2\theta_{\nu}
+\frac{1}{2}\left(
\frac{\mu_{22}}{\gamma_{2}}-\frac{\mu_{11}}{\gamma_{1}}
\right)\sin 2\theta_{\nu},
\\
-\left(\frac{\mu}{\gamma}\right)_{\mu\mu}&=
\frac{\mu_{11}}{\gamma_{1}}\,\sin^2 \theta_{\nu}
+\frac{\mu_{22}}{\gamma_{2}}\,\cos^2 \theta_{\nu}
-\frac{\mu_{12}}{\gamma_{12}}\,\sin 2\theta_{\nu}.
\nonumber
\end{align}
Here $\gamma_1$ and $\gamma_2$ are the Lorenz factors of the massive neutrinos, and
\begin{equation}
\label{eq:GammaDefinition}
\frac{1}{\gamma_{12}}=\frac{1}{2}\left(\frac{1}{\gamma_1}+\frac{1}{\gamma_2}\right).
\end{equation}
The problem of the neutrino propagating in the interstellar space
can be solved precisely for the 
constant interstellar matter density and magnetic field 
(see \cite{kurashvili2017spin} for details).


\section{\label{sec:solution} Solution of the Lindblad equation}

Giant magnetic fields generated by cosmic objects impact neutrinos remotely from the sources of the magnetic fields, i.e., in the interstellar space.  The stochastic magnetic field includes the effects of interstellar fluctuations, galactic winds, cosmic turbulence, and primordial magnetic field fluctuations. The stochastic magnetic fields exert a dissipative effect on the neutrino spin polarization. Therefore the evolution of neutrinos in stochastic fields is not unitary and should be described by the Lindblad master equation. We note that the stochastic component of the field is quite strong.  In particular, in the centre of galaxy M51, the ratio between stochastic and regular components is on the order of 10$\%$ \cite{houde2013characterizing}. 
The Lindbladian approach for neutrinos is described in detail in recent publications \cite{PhysRevD.103.036011,kurashvili2021coherence}.

We prepare the initial state of the system in the mixture of the flavour states:
\begin{eqnarray}\label{mixture of the flavour states}
&&\hat{\varrho}(0)=\mathcal{\hat{I}}=a_1\vert\nu_{e}^{L}\rangle\langle\nu_{e}^{L}\vert+a_2\vert\nu_{\mu}^{L}\rangle\langle\nu_{\mu}^{L}\vert+\nonumber\\
&& a_3\vert\nu_{e}^{R}\rangle\langle\nu_{e}^{R}\vert+a_4\vert\nu_{\mu}^{R}\rangle\langle\nu_{\mu}^{R}\vert, 
\end{eqnarray}
and propagate the initial state in the interstellar medium using the Hamiltonian 
Eq. (\ref{Hamiltonian of the problem}). 

The interstellar magnetic field has two contributions: deterministic part $\vec{B}$ that enters in 
Eq. (\ref{Hamiltonian of the problem}) 
and stochastic magnetic field related to the cosmic dust.
The stochastic field is characterized by the mean value 
$\langle \vec{h}(t)\rangle=0$ and the correlation function can be presented 
in the form 
$\langle h_\alpha(t)h_\beta(0)\rangle=\delta_{\alpha\beta}\eta B^2 f(t)$, 
where $\eta=\langle h^2\rangle/B^2$, and $f(t)$ takes 
the $\delta$-correlator form $f(t)=L_0\delta(t)$ if the correlation length $L_0$ 
is much less than the neutrino oscillation length $L_{osc}$ 
(see \cite{kurashvili2021coherence} and references therein).
This allows us to rewrite the correlation function 
$\langle h_\alpha(t)h_\beta(0)\rangle$ in the following form:
$\langle h_\alpha(t)h_\beta(0)\rangle=
\frac{\delta_{\alpha\beta}W^2}{2\mu_\nu^2}\,\delta(t)$,
where $\mu_\nu$ is a putative value of the neutrino magnetic moment and 
$W^2=2\eta(\mu_\nu B)^2L_0$ is the dissipation parameter (see below). 
For the interstellar case, one has $B\simeq3$~$\mu$G, and $\eta\sim1$ and 
$L_0\sim50$~pc (see \cite{kurashvili2021coherence} and references therein).

We exploit Novikov's theorem \cite{novikov1965functionals}
and consider  non-perturbative exact master equation for the noise-averaged 
density matrix (see \cite{PhysRevLett.117.080402} for a recent literature).
The density matrix of the system $\hat{\varrho}$ obeys the Lindblad master equation 
in the Novikov's form:
\begin{eqnarray}\label{master equation}
&& \frac{d\hat{\varrho}}{dt}=
-i\left[\hat H, \hat \varrho \right] - 
\frac{W^{2}}{2}\left(\hat{\varrho}\hat{V^2}
+\hat{V^2}\hat{\varrho}-2\hat{V}\hat{\varrho}\hat{V}\right).
\end{eqnarray}
The first term in Eq. (\ref{master equation}) describes unitary evolution and 
the second term proportional to $W^2$ is responsible for dissipative effects. 
We note that the master equation
(\ref{master equation}) preserves the trace of the density matrix.  
The dissipator matrix $V$ has the following general form:
\begin{equation}
\label{eqn:Vmatrix}
V_{ik} = \bra i I_1  \otimes v_2 + I_2 \otimes v_1 \ket k ,
\end{equation}
where $v$ is any $2 \times 2$ matrix, and the
subscripts 1, 2 denote the action of a matrix on the
space of the first and second neutrino, respectively.
The matrix $v$ can be expanded into the basis of $2\times 2$ unit matrix and
three Pauli matrices:

\begin{equation}
v = v_0 I + \vec v \cdot  \vec \sigma.
\label{eqn:vexpansion_coherence}
\end{equation}

We analytically solve Eq. (\ref{master equation}) in the 
eigenbasis of the Hamiltonian $\hat{H}_{eff}$. 

\subsection{\label{subsec:diagonalization}Diagonalization of the Hamiltonian}

The density matrix can be written in different representations. 
For the specified problem of the neutrino motion in the magnetic field, 
the most convenient representation is that in the basis of the eigenfunctions 
of the Hamiltonian in Eq. (\ref{Hamiltonian of the problem}). 
In the following, we neglect the neutrino interaction
with the interstellar matter, which is substantially weaker than both
the neutrino interaction with the interstellar magnetic field
and the neutrino vacuum oscillation frequency. 
In the magnetic field Hamiltonian (\ref{eq:H_EM}),
we neglect terms proportional to the longitudinal component of the field due to
the large Lorentz factors of ultrahigh-energy neutrinos.
Taking the mass basis, the resulting effective Hamiltonian can be written as
\begin{equation}
\label{eq:HamiltonianMassRepresentation}
\hat H_{\mathrm{eff}}^{\mathrm m} =
\begin{pmatrix}
-\omega_\nu && \mu_{11} B && 0 && \mu_{12} B \\
\mu_{11} B && -\omega_\nu && \mu_{12} B && 0 \\
0 && \mu_{12} B && \omega_\nu && \mu_{22} B  \\
\mu_{12} B && 0 && \mu_{22} B && \omega_\nu
\end{pmatrix}.
\end{equation}
The effective Hamiltonian can be diagonalized in two consecutive steps.
At first, we combine the states of the same mass and create the symmetric 
and antisymmetric combinations of the spin states:
\begin{equation}
\label{eq:nuBsqrt2}
\nu_{i}^{\pm} = \frac 1 {\sqrt{2}}
\left( \psi_{i+} \pm \psi_{i-}\right), \text { } i=1,2.
\end{equation}
The corresponding transformation matrix reads
\begin{equation}
\label{eq:nuBsqrt2_matr}
T_S = \frac 1 {\sqrt 2}
\begin{pmatrix}
1 && 1 && 0 && 0
\\
1 && -1 && 0 && 0
\\
0 && 0 && 1 && 1
\\
0 && 0 && 1 && -1
\end{pmatrix},
\end{equation}
or, in terms of $2 \times 2$ blocks,
\begin{equation}
\label{eq:nuBsqrt2_matr_2times2}
T_S = \frac 1 {\sqrt 2} 
\begin{pmatrix}
\sigma_3 + \sigma_1 && 0
\\
0 && \sigma_3 + \sigma_1
\end{pmatrix},
\end{equation}
where $\sigma_{1,3}$ are Pauli matrices.
The Hamiltonian takes the form:
\begin{align}
\label{eq:HamiltonianPartiallyRotated}
\begin{pmatrix}
-\omega_\nu + \mu_{11} B && 0 && \mu_{12}  B && 0 
\\
0 && 
-\omega_\nu- \mu_{11} B &&  0 && -\mu_{12} B
 \\
\mu_{12} B && 0  && \omega_\nu+ \mu_{22} B && 0
\\
0 && - \mu_{12} B && 0  && \omega_\nu-  \mu_{22} B
\end{pmatrix}.
\end{align}
For full diagonalization we use the following matrix:
\begin{equation}
\label{eq:FinalDiagonaliation}
T_{b} =
\begin{pmatrix}
c_b && 0 && -s_b&& 0
\\
0 && c_b && 0 && s_b
\\
s_b && 0  && c_b && 0
\\
0 && -s_b && 0 && c_b
\end{pmatrix}=
\begin{pmatrix}
c_b I && -s_b \sigma_3 
\\
s_b \sigma_3 && c_b I
\end{pmatrix},
\end{equation}
where $c_b=\cos\theta_{\mathrm B}$ and $s_b=\sin\theta_{\mathrm B}$ with
\begin{equation}
\label{eq:thetaB}
\theta_{\mathrm B} =
\frac 1 2 \arctan \left[\frac{ \mu_{12}B}{\omega_\nu - 
\frac 1 2 (\mu_{11} - \mu_{22}) B }
\right].
\end{equation}
The eigenvalues of the Hamiltonian are equal to
\begin{align}
\label{eq:SolutionMuFull}
\kappa   = &\pm\frac{(\mu_{11} + \mu_{22}) B} 2  \pm 
\\
\nonumber
&\sqrt{\left (\omega_\nu - \frac {\mu_{11} - \mu_{22}}
2 B\right)^2+ (\mu_{12}B)^2}. 
\end{align}
If we set all magnetic moments equal to the same value 
$\mu_\nu$,
\begin{equation}
\label{eq:EqualityOfMagneticMoments}
\mu_{11} = \mu_{22} = \mu_{12} =\mu_\nu,
\end{equation}
then
\begin{equation}
\label{eq:characteristic_hamiltonian_solution}
\kappa = \pm \left[ \sqrt{ \omega_\nu^2 + (\mu_\nu B)^2} 
\pm \mu_\nu B \right],
\end{equation}
or
\begin{equation}
\label{eq:omegaNomegaB}
\kappa=\frac {\pm(\omega_{\mathrm N} \pm \omega_{\mathrm B})} 2,
\end{equation}
where
\begin{equation}
\label{eq:omega_definitions}
\omega_{\mathrm B} = 2\mu_\nu B, \, 
\omega_{\mathrm N} = 2\sqrt{\omega_\nu^2 + \omega_{\mathrm B}^2}.
\end{equation}
The ``magnetic'' angle with these assumptions is
\begin{equation}
\label{eq:thetaB2}
\theta_{\mathrm B} = \frac 1 2 \arctan \left[\frac {\omega_{\mathrm B}}
{\sqrt{\omega_{\mathrm N}^2 - \omega_{\mathrm B}^2} }\right].
\end{equation}

Eqs. (\ref{eq:HamiltonianPartiallyRotated}) and (\ref{eq:FinalDiagonaliation})
define the transformations from the mass basis to the basis defined by the
Hamiltonian.
The full transform is
\begin{equation}
\label{eq:FullTB}
T_{\mathrm B} = T_b T_S =
\begin{pmatrix}
C_{\mathrm B} && -S_{\mathrm B}
\\
S_{\mathrm B} && C_{\mathrm B}
\end{pmatrix},
\end{equation}
where $C_B$, $S_B$ are $2 \times 2$ matrices defined as
\begin{align}
\label{eq:CB}
C_{\mathrm B} & = \frac {c_b} {\sqrt 2} (\sigma_1 + \sigma_3) = 
\frac {c_b} {\sqrt 2}
\begin{pmatrix}
1 && 1 \\ 1 && -1
\end{pmatrix}, 
\\
\label{eq:SB}
S_{\mathrm B} & = \frac {s_b} {\sqrt 2}  \sigma_3 (\sigma_1 + \sigma_3) = 
\frac{s_b}{\sqrt 2}
\begin{pmatrix}
1 && 1 \\ -1 && 1
\end{pmatrix}.
\end{align}

The density matrix in the basis of the Hamiltonian $H_{\mathrm{eff}}$ is
related to the matrix in the mass and flavour representations by the following
formulas:
\begin{equation}
\label{eq:rhobtofm}
\hat \varrho^{\mathrm B} = T_{\mathrm B} \hat \varrho^{\mathrm m}
T_{\mathrm B}^{-1} = (T_{\mathrm B} T_{\mathrm F}) \hat \varrho^{\mathrm f}
(T_{\mathrm B} T_{\mathrm F})^{-1},
\end{equation}
where $T_{\mathrm F}$ is the operator relating
the density matrices in the flavour and mass representations:
\begin{equation}
\label{eq:masstoflavourbasis}
T_{\mathrm F} = 
\begin{pmatrix} 
c_\nu && 0 && s_\nu  &&  0 \\
0 && c_\nu && 0 && s_\nu \\
-s_\nu && 0 && c_\nu && 0 \\
0 && -s_\nu && 0 && c_\nu
\end{pmatrix},
\end{equation}
where $c_\nu = \cos \theta_\nu$, $s_\nu = \sin \theta_\nu$. 

\subsection{\label{subsec:equation}Lindblad equation}

The Lindblad master equation is written in the eigenbasis of the 
Hamiltonian of the neutrino interaction with the magnetic field:
\begin{align}
\label{eq:lindblad}
\frac{d \hat \varrho^{\mathrm B}_{nm}}{dt} 
=& -i \omega_{nm}\hat \varrho^{\mathrm B}_{mn} - 
\frac{w^2}{2} \sum_q 
\left(\hat \varrho^{\mathrm B}_{nq} V^2_{qm} + 
V^2_{nq} \hat \varrho^{\mathrm B}_{qm}\right)
\nonumber \\
&+ w^2 \sum_{q,s} V_{nq} \hat \varrho^{\mathrm B}_{qs} V_{sm},
\end{align}
where $\omega_{nm} = E_n - E_m$, $n,m = 1,..,4$.
The energy levels of the neutrino in the magnetic field are
expressed in Eq. (\ref{eq:omegaNomegaB}) and
below we adopt the following ordering scheme of energy states:
\begin{equation}
\label{eq:Level_ordering}
E_{1,2} = \frac{-\omega_{\mathrm N} \pm \omega_{\mathrm B}} 2, \mbox{ }
E_{3,4} = \frac{\omega_{\mathrm N} \pm \omega_{\mathrm B}} 2.
\end{equation}

The general form of the matrix $V$ reads:
\begin{equation}
\label{eq:Vmatrix}
V_{ik} = \bra i I_1  \otimes v_2 + I_2 \otimes v_1 \ket k ,
\end{equation}
where $v$ is a $2 \times 2$ matrix defining the interaction with the stochastic
part of the magnetic field and acting separately on subspaces spanned by
basis components with numbers $1$, $2$, and $3$ $4$, respectively.
Due to this form of the interaction operator, the
overall system of $16$ linear equations splits into $4$ different subsystems of
$4$ linear equations for respective quadrants of the full density matrix.

We write the density matrix as a set of $4$ independent quadrants:
\begin{equation}
\label{eq:dmatrix_R}
\hat \varrho^{\mr B} =
\begin{pmatrix}
\varrho^{(11)} & \varrho^{(12)} \\
\varrho^{(21)} & \varrho^{(22)}
\end{pmatrix},
\end{equation}
where $\varrho^{(\alpha)}$ are the $2 \times 2 $ minors of the
density matrix each obeying closed linear equations:
\begin{align}
\label{eq:FourLinearSystems}
\frac{d\varrho^{(ab)}_{ik}}{dt} = 
&i (\varepsilon^{ab}\omega_{\mathrm N} - \varepsilon_{ik}\omega_{\mathrm B})
\varrho^{(ab)}_{ik} -
\\ \nonumber
& \frac {w^2} 2
\sum_l \left(\varrho^{(ab)}_{il}v^2_{lk}+ v^2_{il}\varrho^{(ab)}_{lk}\right) +
\\ \nonumber
&w^2 \sum_{l,m} v_{il}\varrho^{(ab)}_{lm}v_{mk},
\end{align}
with $a, b = 1, 2$ representing the quadrant indices as in
Eq. (\ref{eq:dmatrix_R}), and $i, k=1, 2$
the indices inside these quadrants.
$\epsilon^{ab}$ and $\epsilon_{ik}$ are two-dimensional
symbols antisymmetric with respect to indices.

The term proportional to $\epsilon^{ab}$ is the same for all minor elements 
and therefore
\begin{equation}
\label{eq:IncludeTimeDependence}
\varrho^{(ab)}(t) = e^{i\varepsilon^{ab}\omega_N t}\bar\varrho^{(ab)}(t).
\end{equation}

The matrices $\bar \varrho^{(ab)}$ and
$v$ are expanded into the linear combinations of
the three Pauli matrices and $2 \times 2$ unit matrix:
\begin{equation}
\label{eq:Rsigma}
\bar \varrho^{(ab)} = r^{(ab)}_0 I + \vec r^{(ab)} \cdot \vec \sigma,
\end{equation}
where the coefficients are defined by:
\begin{equation}
\label{eq:Rsigma_trace}
r_i^{(ab)} = \frac 1 2 \mr{Tr}\{ \varrho^{(ab)}\sigma_i \}, \, \,
r_0^{(ab)} = \frac 1 2 \mr{Tr}\{ \varrho^{(ab)} \}.
\end{equation}
The same expansion we exploit for the matrix $v$:
\begin{equation}
\label{eq:vsigma}
v = v_0 I + \vec v \cdot \vec \sigma.
\end{equation}
Using these expansions,
one can show that the terms containing
$v_0$ do not contribute to the dissipation process, since
they are proportional to a unit matrix that commutes with all other
matrices, and the second and third terms of Eq. (\ref{eq:FourLinearSystems})
give contributions equal in magnitude but with opposite signs.
The vector $\vec v$ can be represented as
\begin{equation}
\label{eq:VAlphaBeta}
\vec v = \vert \vec v \vert
( \cos \varphi \sin \beta, \sin \varphi \sin \beta , \cos \beta),
\end{equation}
choosing $\varphi=0$ to make it purely real,
and normalizing to unity, which can  be done by a proper rescaling of the
coefficient $w^2$, one deduces
\begin{equation}
\label{eq:VBeta}
\vec v = (\sin \beta, 0 , \cos \beta).
\end{equation}
The angle parameter $0<\beta < \pi/2$ defines strength of the dissipation.
The dissipation is zero if $\beta=0$ and reaches
its maximum value for $\beta=\pi/2$.
In the present work we set $\beta = \pi/4$.
After redefining time,
\begin{equation}
\label{eq:timeredefine}
\tau =  2 w^2  t,
\end{equation}
and frequencies,
\begin{equation}
\label{eq:frequencyredefine}
\bar\omega_{\mathrm {B,N}} = \frac{\omega_{\mathrm{B,N}} } {2w^2},
\end{equation}
and using above expansions, through lengthy and rather involved calculations,
we deduce the following system of equations for the coefficients $r_i$ of
each quadrant
\begin{align}
\label{eqn:diff_r0_alt}
\frac d {d\tau} r_0(\tau) & = 0,
\\
\label{eqn:diff_r1_bar}
\frac d {d\tau} r_1 (\tau)& = -c_\beta^2  r_1(\tau)  +  \bar \omega_{\mathrm B}
r_2(\tau)  +  c_\beta s_\beta r_3(\tau),
\\
\label{eqn:diff_r2_alt}
\frac d {d\tau} r_2(\tau) & = - \bar \omega_{\mathrm B} r_1(\tau) 
-c_\beta^2 r_2(\tau),
\\
\label{eqn:diff_r3_alt}
\frac d {d\tau} r_3(\tau) & =  c_\beta s_\beta r_1(\tau) -s_\beta^2 r_3(\tau),
\end{align}
where the notations $c_\beta = \cos \beta$, $s_\beta = \sin \beta$ were used.

For solving the linear system we diagonalize the matrix:
\begin{equation}
\label{eq:MR1}
\mathcal{M} =
\begin{pmatrix}
- c_\beta^2 && \homega_{\mathrm B} && s_\beta \cdot c_\beta
\\
-\homega_{\mathrm B} && - c_\beta^2  && 0
\\
s_\beta \cdot c_\beta  && 0 && -s_\beta^2
\end{pmatrix},
\end{equation}
and find roots $\{\nu_k\}$  of the cubic polynomial equation.
The solution can be represented in the form:
\begin{equation}
\label{eq:solution_R1}
r_i (\tau) =  \sum_{k=1}^{3} C_{ik} e^{i\nu_k \tau},
\end{equation}
where the integration constants are given by 
\begin{align}
\label{eq:R1_c1}
C_{i1} & = \frac {B_{0i} \nu_2 \nu_3 - B_{1i} (\nu_2 + \nu_3)
+ B_{2i}}
{(\nu_1 - \nu_2)(\nu_1 - \nu_3)},\\
\label{eq:R1_c2}
C_{i2} & = \frac{B_{0i} \nu_1 \nu_3 - B_{1i} (\nu_1 +\nu_3)
+B_{2i}}
{(\nu_2 - \nu_1)(\nu_2 - \nu_3)},\\
\label{eq:R1_c3}
C_{i3} & = \frac{B_{0i} \nu_1 \nu_2 - B_{1i} (\nu_1 + \nu_2)
+B_{2i}}
{(\nu_3 - \nu_1)(\nu_3 -\nu_2)},
\end{align}
where
\begin{align}
\label{eq:R1betas}
B_{0i} & = r_i(0), \, \,
\\
B_{1i} & = \sum_k \mathcal{M}_{ik}r_{k}(0), \, \,
\\
B_{2i} & = \sum_{kl} \mathcal{M}_{ik}
\mathcal{M}_{kl}r_{l}(0).
\end{align}


\section{\label{sec:measures}Entropic measures}

After involved calculations we obtain analytical results 
for conditional von Neumann entropies. The full spin-flavour density matrix can be written in the
corresponding basis in the form:
\begin{equation}
\label{eq:MatrixRhoSigmaNu}
\hat \varrho = \sum_{\nu \nu' \sigma \sigma'}
\rho^{(\nu \nu')}_{\sigma \sigma'} \ket \nu \ket \sigma \bra \nu' \bra \sigma'.
\end{equation}
To obtain the reduced flavour density matrix, we take the trace over the spin states
\begin{equation}
\label{eq:MatrixReduced_nu}
\hat \varrho_{\nu }  = 
\sum_{\sigma} \bra{\sigma} \hat \varrho \ket{\sigma} =
\sum_{\nu \nu'}\sum_\sigma \rho^{(\nu\nu')}_{\sigma\sigma}\ket{\nu} \bra{\nu'}.
\end{equation}

The reduced density matrix in the first line of 
Eq. (\ref{Quantum memory1})
is obtained by projecting the matrix on the eigenstates of
$\sigma_z$, $\ket 1$ and $\ket 0$ (or $\psi_{1,2}$).
The result is expressed in terms of
the $4 \times 4$ matrices $\rho^{(\nu\nu')}_{\sigma\sigma}$
\begin{equation}
\label{eq:MatrixReduced_flavour}
\hat \varrho_{\nu} = 
\begin{pmatrix}
\rho^{(ee)}_{11} + \rho^{(ee)}_{00} && 
\rho^{(e\mu)}_{11} + \rho^{(e\mu)}_{00} 
\\
\rho^{(\mu e)}_{11} + \rho^{(\mu e)}_{00} && 
\rho^{(\mu \mu)}_{11} + \rho^{(\mu \mu)}_{00} 
\end{pmatrix}.
\end{equation}
The eigenvalues of $\hat \varrho_{\nu}$ are
\begin{equation}
\label{eq:Rhonu_eigenvalues}
\lambda^\nu_{1,2} = \frac {(\varrho^{(ee)}_\nu + \varrho^{(\mu\mu)}_\nu) \pm 
\sqrt{(\varrho^{(ee)}_{\nu} - \varrho_{\nu}^{(\mu\mu)})^2 
+ 4|\varrho_{\nu}^{(e\mu)}|^2 }} 2,
\end{equation}
where for all pairs $(\nu\nu')$,
$\varrho^{(\nu\nu')}_\nu = \varrho^{(\nu\nu')}_{11} + \varrho^{(\nu\nu')}_{00}$.
The reduced spin density matrix is obtained after tracing 
flavour states:
\begin{equation}
\label{eq:MatrixReduced_sigma}
\hat \varrho_{\sigma }  = 
\sum_{\nu} \bra{\nu} \hat \varrho \ket{\nu} =
\sum_{\nu }\sum_{\sigma \sigma} 
\rho^{(\nu\nu)}_{\sigma\sigma'}\ket{\sigma} \bra{\sigma'}.
\end{equation}
We rewrite $\hat \varrho_{\sigma }$ as a $ 2\times 2$ matrix in the following form:
\begin{equation}
\label{eq:MatrixReduced_spin}
\hat \varrho_{\sigma} = 
\begin{pmatrix}
\rho^{(ee)}_{11} + \rho^{(\mu\mu)}_{11} && 
\rho^{(ee)}_{10} + \rho^{(\mu\mu)}_{10} 
\\
\rho^{(ee)}_{01} + \rho^{(\mu \mu)}_{01} && 
\rho^{(ee)}_{00} + \rho^{(\mu \mu)}_{00} 
\end{pmatrix}.
\end{equation}
Eigenvalues $\lambda^{\sigma}_{1,2}$ of Eq. (\ref{eq:MatrixReduced_spin})
have the same structure as in
Eq. (\ref{eq:Rhonu_eigenvalues}) under replacing the elements of the flavour matrix
$\hat \varrho_\nu$ with the corresponding components of $\hat \varrho_\sigma$.

The mutual information between the spin and the flavour is expressed through
the eigenvalues of the spin, flavour and full density matrices:
\begin{align}
\label{eq:SpinFlavorMutualInformation}
I(\nu, \sigma) = -\sum_{i=1}^2 (\lambda^{\nu}_i\log{\lambda^{\nu}_i}
+\lambda^{\sigma}_i \log{\lambda^{\sigma}_i}) +
\\ \nonumber
\sum_{k=1}^4 \lambda_k \log{\lambda_k},
\end{align}
where $\lambda_i$, $i=1,..,4$ denote eigenvalues of the full 
flavour-spin density matrix $\hat\varrho$.

The matrix $\hat \varrho_{R\nu}$ has the following explicit form:
\begin{align}
\label{eq:Rho_R_nu}
\hat \varrho_{R\nu} = &
\sum_{\nu \nu'}\sum_{\sigma} \varrho^{(\nu\nu')}_{\sigma \sigma}
\ket {\sigma} \ket {\nu}\bra{\nu'}\bra{\sigma} =
\\ \nonumber
&\sum_{\nu\nu'} \ket{\nu}\bra{\nu'}
\left( \varrho^{(\nu\nu')}_{11}\ket 1 \bra 1 + 
\varrho^{(\nu\nu')}_{00} \ket 0 \bra 0\right),
\end{align}
and four eigenvalues:
\begin{align}
\label{eq:R1_eigenvalues_12}
\lambda^{R\nu}_{1,2}& = \frac { 
(\varrho^{(ee)}_{11} + \varrho^{(\mu\mu)}_{11}) 
\pm \sqrt{(\varrho^{(ee)}_{11}-\varrho^{(\mu\mu)}_{11})^2 
+ 4 |\varrho^{(e\mu)}_{11}|^2}} 2 ,
\\
\label{eq:R1_eigenvalues_34}
\lambda^{R\nu}_{3,4}& = \frac { 
(\varrho^{(ee)}_{00} + \varrho^{(\mu\mu)}_{00}) 
\pm \sqrt{(\varrho^{(ee)}_{00}-\varrho^{(\mu\mu)}_{00})^2 
+ 4 |\varrho^{(e\mu)}_{00}|^2}} 2 .
\end{align}
After projecting on the basis states 
$\ket{\pm} = 1/\sqrt{2} (\ket 1 \pm \ket 0)$ (or $\phi_{1,2}$),
the second matrix in Eq. (\ref{Quantum memory1}) takes the form:
\begin{align}
\label{eq:Rho_Q_nu_pmbasis}
\hat \varrho_{Q\nu} = 
\sum_{\nu\nu'} \ket{\nu}\bra{\nu'}
\left( \varrho^{(\nu\nu')}_{++}\ket + \bra + + 
\varrho^{(\nu\nu')}_{--} \ket - \bra -\right),
\end{align}
or alternatively
\begin{align}
\label{eq:Rho_Q_nu}
\hat \varrho_{Q\nu} = & \sum_{\nu\nu'} \ket \nu \bra {\nu'}
\frac { \varrho^{(\nu\nu')}_{11} + \varrho^{(\nu\nu')}_{00}} 2
\left(\ket 1 \bra 1 + \ket 0 \bra 0 \right) +
\\ \nonumber	
& \sum_{\nu\nu'}\ket \nu \bra {\nu'}
\frac{\varrho^{(\nu\nu')}_{10} + \varrho^{(\nu\nu')}_{01}} 2
\left( \ket 1 \bra 0 + \ket 0 \bra 1 \right) .
\end{align}
The coefficients in Eqs. (\ref{eq:Rho_Q_nu_pmbasis}) and (\ref{eq:Rho_Q_nu})
are related as
\begin{align}
\label{eq:rho_pp}
\rho^{(\nu\nu')}_{++}&  = \frac 1 2 
\left(\rho^{(\nu\nu')}_{11} + \rho^{(\nu\nu')}_{00}
+ \rho^{(\nu\nu')}_{10} + \rho^{(\nu\nu')}_{01}\right),
\\
\label{eq:rho_mm}
\rho^{(\nu\nu')}_{--}&  = \frac 1 2
\left(\rho^{(\nu\nu')}_{11} + \rho^{(\nu\nu')}_{00}
- \rho^{(\nu\nu')}_{10} - \rho^{(\nu\nu')}_{01}\right).
\end{align}
The matrix (\ref{eq:Rho_Q_nu}) has exactly the same structure as 
$\hat \varrho_{R\nu}$ in the basis $\ket 0$, $\ket 1$, therefore
the expressions for the eigenvalues have the form similar to
Eqs. (\ref{eq:R1_eigenvalues_12}) and (\ref{eq:R1_eigenvalues_34}):
\begin{align}
\label{eq:Q1_eigenvalues_12}
\lambda^{Q\nu}_{1,2}& = \frac { 
(\varrho^{(ee)}_{++} + \varrho^{(\mu\mu)}_{++}) 
\pm \sqrt{(\varrho^{(ee)}_{++}-\varrho^{(\mu\mu)}_{++})^2 
+ 4 |\varrho^{(e\mu)}_{++}|^2}} 2 ,
\\
\label{eq:Q1_eigenvalues_34}
\lambda^{Q\nu}_{3,4}& = \frac { 
(\varrho^{(ee)}_{--} + \varrho^{(\mu\mu)}_{--}) 
\pm \sqrt{(\varrho^{(ee)}_{--}-\varrho^{(\mu\mu)}_{--})^2 
+ 4 |\varrho^{(e\mu)}_{--}|^2}} 2 .
\end{align}

The conditional entropy from the right-hand side of 
Eq. (\ref{correction to the The Kraus}) is
\begin{equation}
\label{eq:Entropy_Conditional_Spin_Flavor}
S(\sigma \vert \nu) = S(\hat \varrho_{\sigma \nu}) - S(\hat \varrho_\nu),
\end{equation}
and mutual information:
\begin{equation}
\label{eq:Mutual_Information_Spin_Flavor}
I(\sigma, \nu) = S(\hat \varrho_\sigma) - S(\sigma \vert \nu).
\end{equation}
We calculate the mutual informations $I(X, \nu)$ which is required
to evaluate the classical correlation according to
Eq. (\ref{The classical correlation term}).
At first we consider $X=\sigma_z$, then
the resulting spin-flavour density matrix is written as
\begin{equation}
\label{eq:Znu_matrix}
\hat \varrho_{\sigma_z \nu} =
\begin{pmatrix}
\rho_{11}^{(\nu\nu')} && 0 \\ 0 && \rho_{00}^{(\nu\nu')}
\end{pmatrix} = \hat \varrho_{R\nu},
\end{equation}
and the reduced spin matrix:
\begin{align}
\label{eq:Z_matrix}
\hat \varrho_{\sigma_z} = 
 \begin{pmatrix}
\rho_{11}^{(ee)} +\rho_{11}^{(\mu\mu)} && 0 \\
0 && \rho_{00}^{(ee)} +\rho_{00}^{(\mu\mu)} 
\end{pmatrix}.
\end{align}
Similarly $X=\sigma_x$,
\begin{equation}
\label{eq:Xnu_matrix}
\hat \varrho_{\sigma_x \nu} = \frac 1 2
\begin{pmatrix}
\rho_{11}^{(\nu\nu')} +\rho_{00}^{(\nu\nu')} &&
\rho_{10}^{(\nu\nu')} +\rho_{01}^{(\nu\nu')} \\
\rho_{10}^{(\nu\nu')} +\rho_{01}^{(\nu\nu')} &&
\rho_{11}^{(\nu\nu')} +\rho_{00}^{(\nu\nu')} 
\end{pmatrix} =
\hat \varrho_{Q\nu}, 
\end{equation}
and
\begin{align}
\label{eq:X_matrix}
\hat \varrho'_{\sigma_x} = 
 \begin{pmatrix}
\rho_{++}^{(ee)} +\rho_{++}^{(\mu\mu)} && 0 \\
0 && \rho_{--}^{(ee)} +\rho_{--}^{(\mu\mu)} 
\end{pmatrix}
\end{align}
in its eigenbasis.

The corresponding mutual informations are:
\begin{align}
\label{eq:Mutual_Information_Spin_R_nu}
I(\sigma_z, \nu) & = S(\hat \varrho_{\sigma_z})-  S(R\vert \nu),
\\
\label{eq:Mutual_Information_Spin_R_nu2}
I(\sigma_x, \nu) & = S(\hat \varrho_{\sigma_x}) - S(Q\vert \nu).
\end{align}

The matrix elements at
the initial moment of time in our notations are expressed as
\begin{align}
\label{eq:DensityMatrixInitial}
&\varrho_{11}^{(ee)} (0) = a_3 = \lambda_1, \,
\varrho_{00}^{(ee)}(0) = a_1 = \lambda_2,
\\
\label{eq:DensityMatrixInitial2}
&\varrho_{11}^{(\mu\mu)}(0) = a_4 = \lambda_3, \,
\varrho_{00}^{(\mu\mu)} (0) = a_2 = \lambda_4,
\end{align}
with all non-diagonal elements equal to zero.
Therefore, at $t=0$
\begin{align}
\label{eq:Rho_eigenvalues_Initial_nu}
&\lambda^\nu_1 (0) = \lambda_1 + \lambda_2, \, 
\lambda^\nu_2 (0)  = \lambda_3 + \lambda_4,
\\
\label{eq:Rho_eigenvalues_Initial_sigma}
&\lambda^\sigma_1 (0)  =  \lambda_1 + \lambda_3, \, 
\lambda^\sigma_2(0) = \lambda_2 + \lambda_4,
\\
\label{eq:Rho_eigenvalues_Initial_sigmaz}
&\lambda^{\sigma_z}_1 (0)  = \lambda_1 + \lambda_3, 
\, \lambda^{\sigma_z}_2(0) = \lambda_2 + \lambda_4,
\\
\label{eq:Rho_eigenvalues_Initial_sigmax}
&\lambda^{\sigma_x}_1(0)  = \frac 1 2, \, \lambda^{\sigma_x}_2(0) = \frac 1 2,
\\
\nonumber 
&\lambda^{R\nu}_1 (0)  = \lambda_1, \, \lambda^{R\nu}_2 (0) = \lambda_2,
\\
\label{eq:R1_eigenvalues_34_Initial}
&\lambda^{R\nu}_3 (0)  = \lambda_3, \, \lambda^{R\nu}_4 (0) = \lambda_4,
\\
\nonumber 
&\lambda^{Q\nu}_{1} (0) = \lambda^{Q\nu}_{2}(0)  =
\frac {\lambda^\nu_{1} (0)} 2 = \frac{\lambda_1 + \lambda_2} 2,
\\
\label{eq:Q1_eigenvalues_24_Initial}
&\lambda^{Q\nu}_{3}(0)  = \lambda^{Q\nu}_{4}(0)  = 
\frac{ \lambda^\nu_{2}(0)} 2 = \frac {\lambda_3 + \lambda_4} 2.
\end{align}
We write down the entropies at the moment of time $t=0$. 
Below $h(x)$ denotes the binomial entropy
$h(x) = -x\log{x} -(1-x)\log{(1-x)}$, where $x$ takes different values
from $0$ to $1$.
The arguments used below are
\begin{align}
\label{eq:Pnu}
p_\nu = \lambda_1 + \lambda_2,
\\
\label{eq:Psigma}
p_\sigma = \lambda_1 + \lambda_3,
\end{align}
which correspond to the fractions of electron neutrinos and
right-handed neutrinos, respectively.
We denote
\begin{equation}
\label{eq:S0}
S_0 = - \sum_i a_i \log{a_i}.
\end{equation}
The entropies are expressed as
\begin{align}
\label{eq:Srho_nusigma}
& S(\hat \varrho_{\nu \sigma}) (0) =  S_0,
\\
\label{eq:Srho_nu}
& S(\hat \varrho_{\nu}) (0) = h(p_\nu), 
\\
\label{eq:Srho_sigma}
& S(\hat \varrho_{\sigma}) (0) = h(p_\sigma),
\\
\label{eq:Srho_Rnu}
& S(\hat \varrho_{R\nu}) (0) =  S_0
\\
\label{eq:Srho_Qnu}
& S(\hat \varrho_{Q\nu}) (0) = 1 + h(p_\nu). 
\\
\label{eq:Srho_nuz}
& S(\hat \varrho_{\sigma_z}) (0) = h(p_\sigma), 
\\
\label{eq:Srho_nux}
& S(\hat \varrho_{\sigma_x}) (0) = 1. 
\end{align}
The conditional entropies are
\begin{align}
\label{eq:S_sigma_nu_t0}
S(\sigma \vert \nu)(0) &=  S_0 - h(p_\nu),
\\
\label{eq:S_Rnu_nu_t0}
S(R \vert \nu) (0) &=  S_0  - h(p_\nu),
\\
\label{eq:S_Qnu_nu_t0}
S(Q \vert \nu) (0) &= 1.
\end{align}
The mutual informations are
\begin{align}
\label{eq:I_sigma_nu_t0}
I(\sigma, \nu)(0) \, \, & = h(p_\sigma) + h(p_\nu) - S_0,
\\
\label{eq:I_sigmaz_nu_t0}
I(\sigma_z, \nu)(0) & = h(p_\sigma) + h(p_\nu) - S_0,
\\
\label{eq:I_sigmax_nu_t0}
I(\sigma_x, \nu)(0) & = 0.
\end{align}

Since the quantum information $I(\sigma_x, \nu)$ is zero,
which is the minimum possible value, the classical correlation
\begin{equation}
\label{eq:J_nu_t0}
J_\sigma(\hat \varrho_{\nu \sigma})(0) = 
I(\sigma_z, \nu)(0),
\end{equation}
and the discord is zero because of Eqs. (\ref{eq:I_sigma_nu_t0})
and (\ref{eq:I_sigmaz_nu_t0}).
It is easy to see, that for a particular choice of the state
Eq. (\ref{mixture of the flavour states}), the last term
in Eq. (\ref{correction to the The Kraus}) is always equal to zero.
Therefore, taking into account that 
$S(R\vert \nu)(0) = S(\sigma\vert \nu)(0)$,
the entropic uncertainty relation is satisfied at $t=0$:
\begin{equation}
\label{eq:Kraus_t0}
S(Q\vert \nu)(0) = -2 \log c(P, Q) = 1.
\end{equation}


\section{\label{sec:results}Results}
In what follows, we consider the density matrix
for the ultrahigh-energy neutrino in an interstellar magnetic field
obtained through 
the Lindblad equation and the time-dependent uncertainty relations.
As in \cite{kurashvili2017spin} we use putative magnetic moments 
$\mu_\nu=2.6\times 10^{-12} \mu_{B}$ and $\mu_\nu = 4.5 \times 10^{-12}\mu_B$
with Bohr magneton $\mu_B = 5.79\times 10^{-9}$ eV/G,
which correspond to the upper experimental limit on the dipole
magnetic moment of neutrinos obtained from observations of red
giants in the globular clusters \cite{Viaux:2013hca}.
Thus the energy of spin-field interaction is $\mu_\nu B \sim 10^{-25}$ eV.
The square mass difference is based on solar neutrino
observation 
$\Delta m^2 = \Delta m_\mathrm{solar}=7.37 \times 10^{-5}$ eV$^2$,
and the vacuum mixing angle is $\sin^2 \theta_\nu = 0.297$
\cite{ParticleDataGroup:2020ssz}.
As in the previous work we focus on ultrahigh-energy neutrinos
$E_\nu \sim 10^{20}$ eV consistent
with Greisen-Zatsepin-Kuz’min (GZK) cut-off for cosmic particles energy
\cite{Greisen:1966jv,Zatsepin:1966jv}.
The oscillation energy of the neutrino in this case is of the 
same order as the interaction with the magnetic field.
To illustrate the interplay between the vacuum and magnetic oscillations,
 we put the vacuum oscillation and magnetic interaction energies to
satisfy $\omega_\mathrm{V}/\omega_\mathrm{B}=5$.
The stochastic part of the magnetic field $w^2$ is assumed to be
equal to 10\% of the total value of $B$ as it is discussed
in \cite{PhysRevD.103.036011} and \cite{kurashvili2021coherence}.
The results are obtained in dimensionless time and energy units of 
Eq. (\ref{eq:timeredefine}), 
$1/(2w^2)$ corresponding to the unit on the physical time scale,
and all frequencies are normalized as in Eq. (\ref{eq:frequencyredefine}).
The dissipation angle is set equal to $\beta = \pi/4$.

\begin{figure}[h!]
\centering
\includegraphics[width=0.45\textwidth]{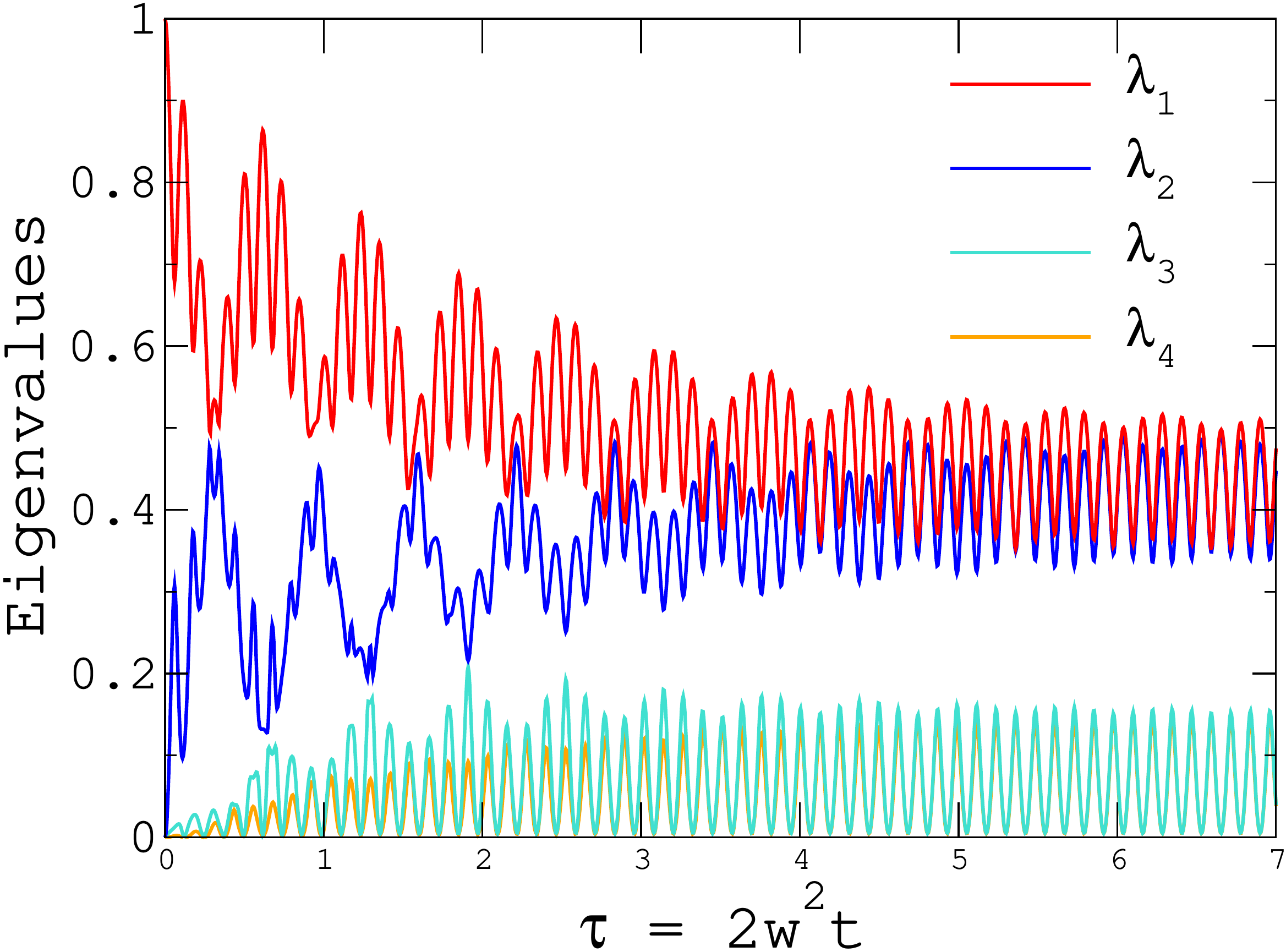}
\caption{\label{fig:Lambdas1}
Eigenvalues of the full density matrix for the choice of initial
condition $\hat \varrho^{\mathrm f}(\tau=0)
= \mathrm{diag}[1, 0, 0, 0]$ as in Eq. (\ref{eq:InitialCondition1}).
The time scale at the horizontal axis
is taken in dimensionless units found by Eq. (\ref{eq:timeredefine}).
The corresponding dimensionless frequencies 
defined as in Eq. (\ref{eq:frequencyredefine}) 
for neutrino vacuum oscillations and spin-magnetic field interaction
are $\homega_{\mathrm V} =5$ and $\homega_{\mathrm B}=1$, correspondingly.
The stochastic component constitutes 10\% of the total 
magnetic field.
The angle parameter is $\beta = \pi/4$ (see Eq. (\ref{eq:VBeta})).
In real units $\omega_B = \mu_\nu B$, where the interstellar magnetic field is
$B = 2.93$ $\mu$G, and $\mu_\nu=2.6\times 10^{-12} \mu_{B}$, with the Bohr
magneton $\mu_B = 5.79 \times 10^{-9}$ eV/G.
The energy of the spin-field  interaction is of the order of $10^{-25}$ eV.
The energy of neutrino is of the order $E_\nu \sim 10^{20}$ eV, and the energy 
of its
interaction with the matter is of the order of $10^{-30}$ eV, 
which can be safely ignored.
}
\end{figure}

\begin{figure}[h]
\centering
\includegraphics[width=0.45\textwidth]{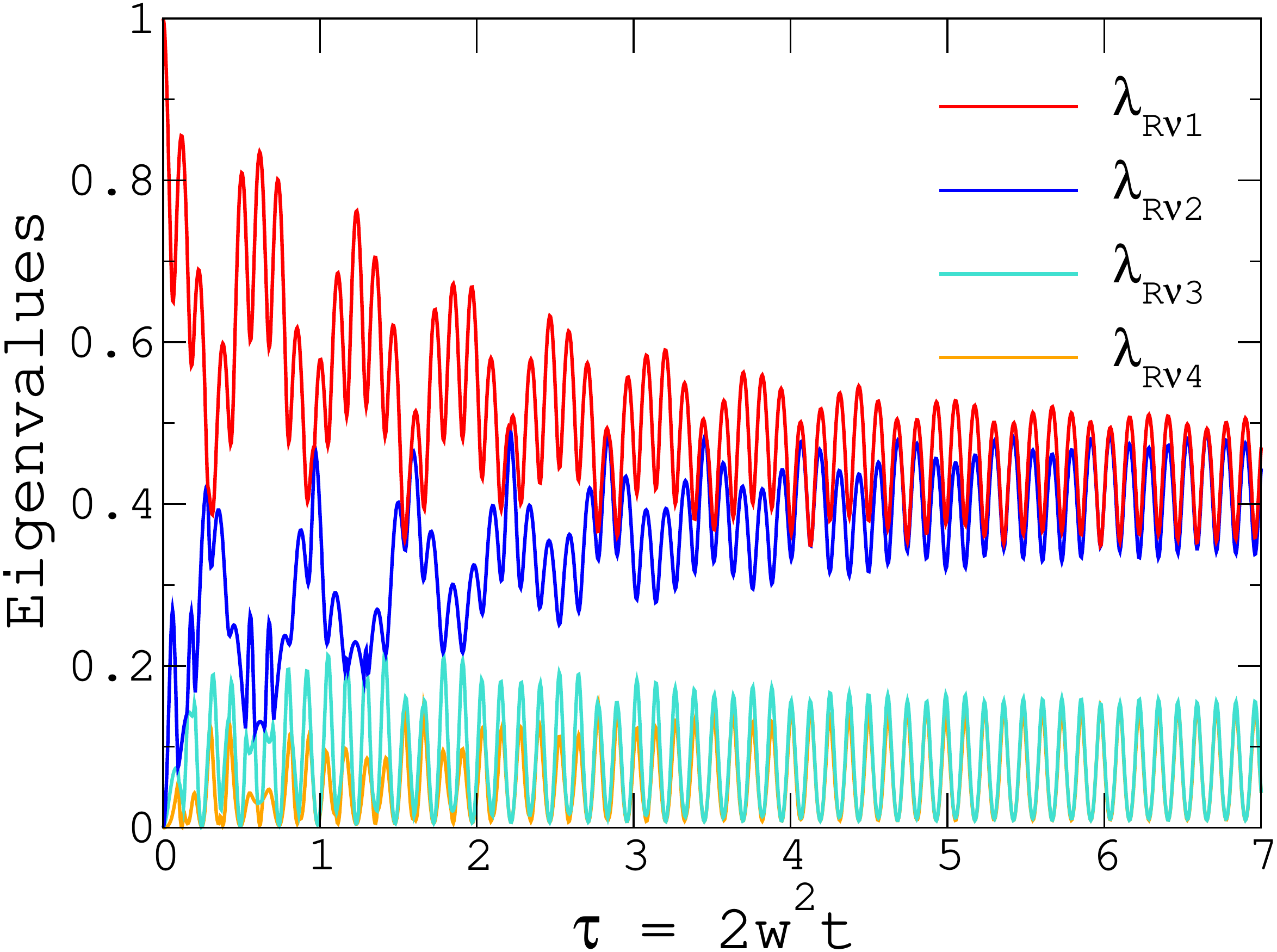}
\caption{
\label{fig:LambdasRnu1}
Eigenvalues of the matrix $\hat \varrho_{R\nu}$  vs $\tau$
in the same conditions
as in Fig. \ref{fig:Lambdas1}.}
\end{figure}

\begin{figure}[h]
\centering
\includegraphics[width=0.45\textwidth]{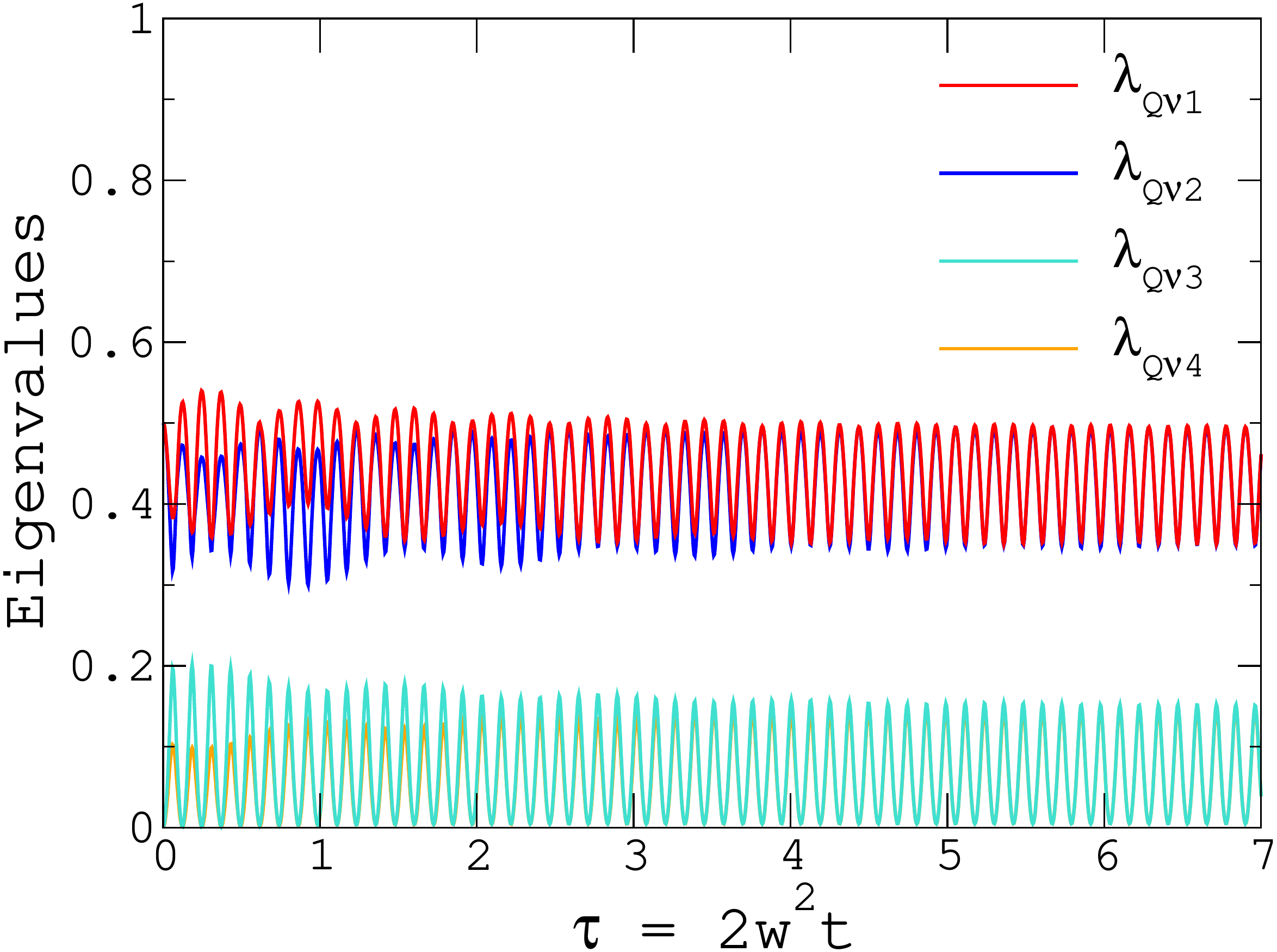}
\caption{
\label{fig:LambdasQnu1}
Eigenvalues of the matrix $\hat \varrho_{Q\nu}$ vs $\tau$ in the same conditions 
as in Fig. \ref{fig:Lambdas1}.}
\end{figure}

\begin{figure}[h!]
\centering
\includegraphics[width=0.45\textwidth]{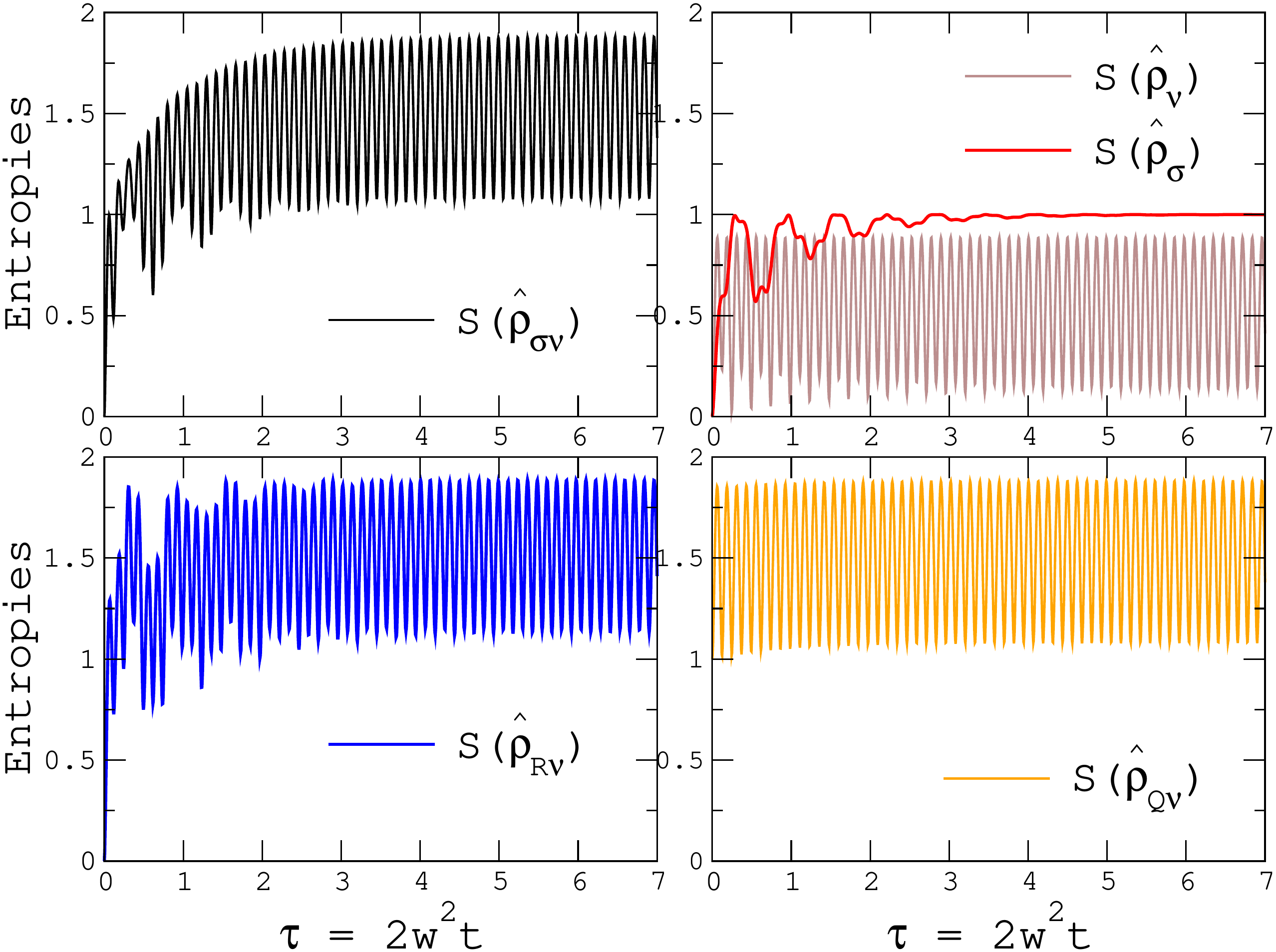}
\caption{
\label{fig:4Entropies1}
Dependence of the entropies corresponding to the full
($\hat \varrho_{\sigma\nu}$, $\hat \varrho_{R\nu}$, $\hat \varrho_{Q\nu}$)
and reduced ($\hat \varrho_\nu$, $\hat \varrho_{\sigma}$) matrices on
the time parameter $\tau$
for the choice of initial
condition as in Eq. (\ref{eq:InitialCondition1}).
Top left corner: the full entropy calculated on the basis of eigenvalues
of the full density matrix $\hat \varrho_{\sigma\nu}$
expressed by Eq. (\ref{eq:MatrixRhoSigmaNu})
shown in Figure \ref{fig:Lambdas1}. 
The value at $\tau=0$ is $0$ and rises as the system thermalizes.
Top right corner: entropies of reduced density matrices 
$\hat \varrho_\nu$ and $\hat \varrho_\sigma$ expressed by
Eqs. (\ref{eq:MatrixReduced_nu}) and (\ref{eq:MatrixReduced_sigma}),
respectively.
Bottom: entropies of $\hat \varrho_{R\nu}$ (left) and 
$\hat \varrho_{Q\nu}$ (right)
calculated with eigenvalues shown in 
Figs. \ref{fig:LambdasRnu1} and \ref{fig:LambdasQnu1}.
}
\end{figure}

In the initial state at $t=0$ only the right-handed electron neutrinos
are present (neutrino flux after passing through a spin filter as it 
is discussed in \cite{PhysRevD.103.036011}):
\begin{equation}
\label{eq:InitialCondition1}
\hat \varrho(0) = \ket {\nu_e^R} \bra {\nu_e^R}.
\end{equation}
The evolution of the eigenvalues of the full density matrix 
is shown in Fig. \ref{fig:Lambdas1}.
The eigenvalues of the
projected matrices $\hat \varrho_{R\nu}$ and $\hat \varrho_{Q\nu}$
are presented in Figs. \ref{fig:LambdasRnu1} and \ref{fig:LambdasQnu1}.
We see that after evolving through the Lindbladian channel, the density matrix relaxes, and we observe steady-state oscillations. 
We plot the entropies required for the calculation of the uncertainty relation.
Fig. \ref{fig:4Entropies1} shows the full entropy, as well as the
entropies corresponding to the different projected density matrices
for the choice of the initial conditions as in 
Eq. (\ref{eq:InitialCondition1}). 
While other entropies exhibit steady-state oscillations with significant 
amplitude, the reduced density matrix of the spin subsystem becomes constant. 
This fact shows that spin thermalizes entirely while the flavour 
subsystem still possesses coherence.  
The reason is that the thermal bath, i.e., random magnetic field, is 
directly coupled to the spin but not to the flavour subsystem. 
The flavour subsystem experiences the thermalization
process only through the spin-flavour oscillation process.

\begin{figure}[h!]
\centering
\includegraphics[width=0.45\textwidth]{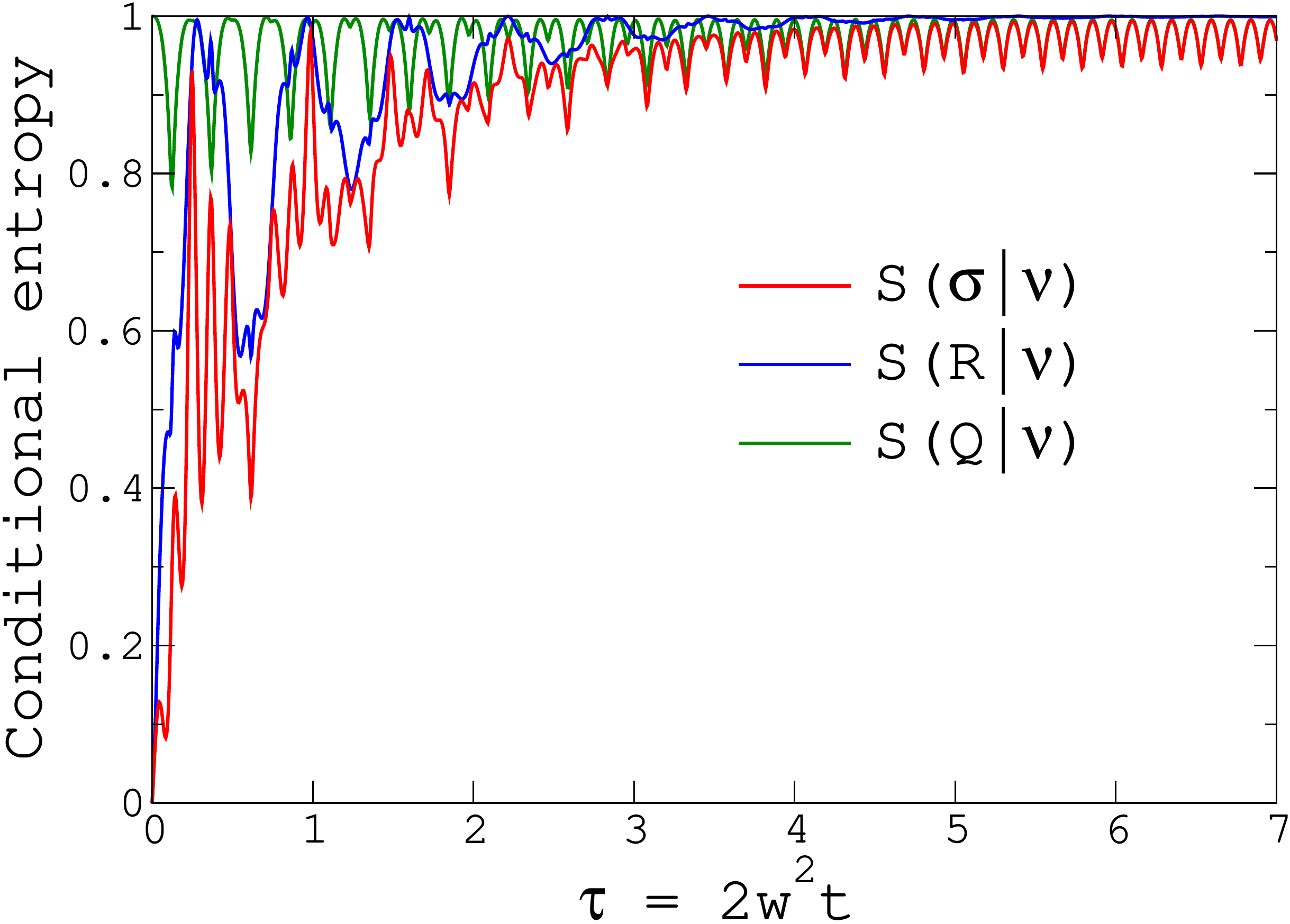}
\caption{
\label{fig:Conditional1}
Conditional entropies $S(\sigma \vert \nu)$, $S(R\vert \nu)$,
and $S(Q\vert \nu)$ as functions of the time parameter $\tau$.
The entropies are calculated based on the entropies
$S(\hat \varrho_{\sigma \nu})$, $S(\hat \varrho_{R\nu})$, 
$S(\hat \varrho_{Q\nu})$  and $S(\hat \varrho_\nu)$ shown in 
Figure \ref{fig:4Entropies1}.
}
\end{figure}

The conditional entropies involved in the entropic uncertainty relation,
are presented in Fig. \ref{fig:Conditional1}. We see persistent small amplitude oscillations in the steady-state regime.

\begin{figure}[h!]
\centering
\includegraphics[width=0.45\textwidth]{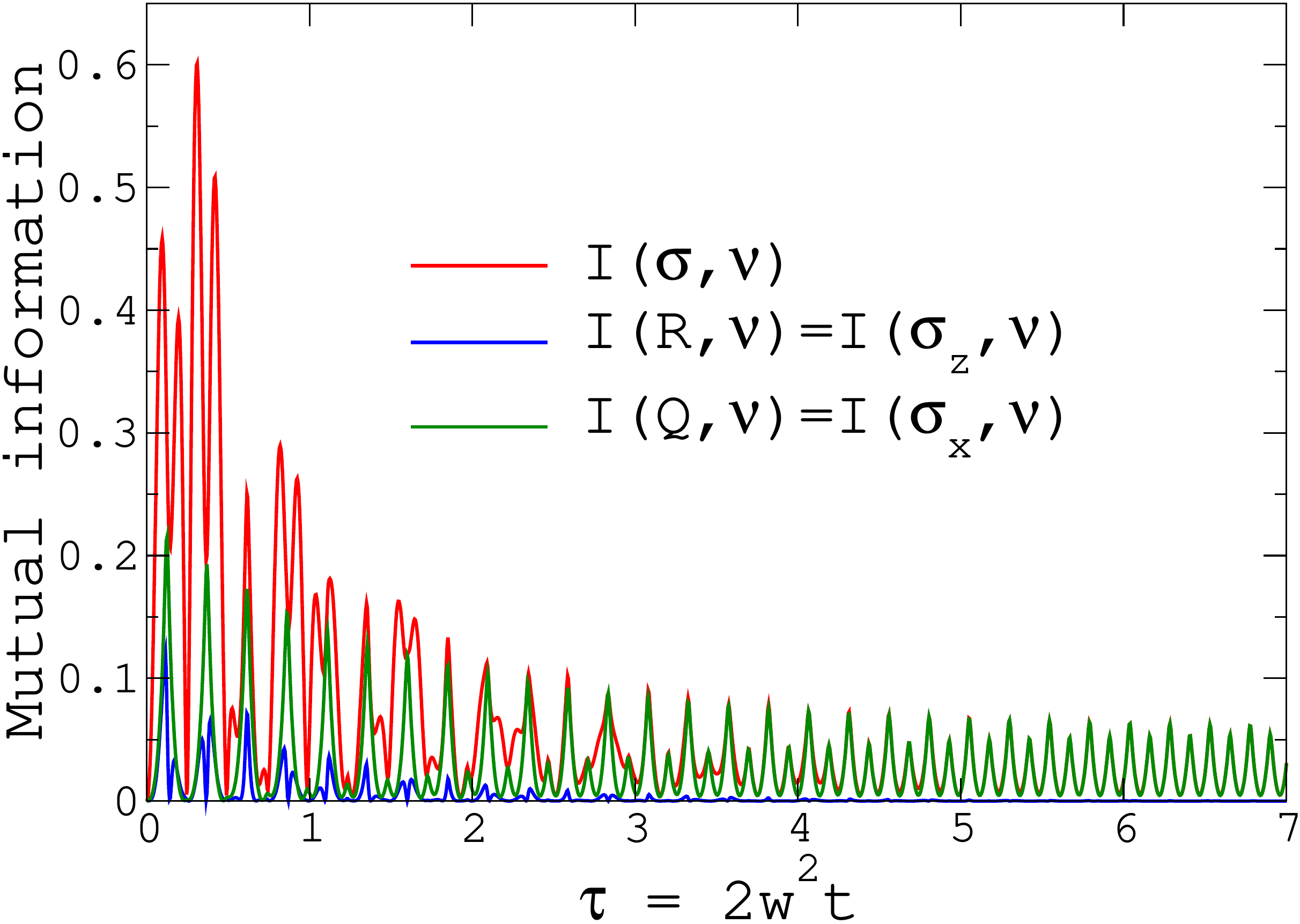}
\caption{
\label{fig:Informations1}
$\tau$-dependence of 
the mutual informations 
defined by Eqs. (\ref{eq:Mutual_Information_Spin_Flavor}),
(\ref{eq:Mutual_Information_Spin_R_nu}) and 
(\ref{eq:Mutual_Information_Spin_R_nu2})
for the choice of initial
conditions as in Eqs. (\ref{eq:InitialCondition1}).
The equalities follow from the identity 
of corresponding density matrices due to the specifics of our problem, as it is
shown in Eqs. (\ref{eq:Znu_matrix}) and (\ref{eq:Xnu_matrix}).}
\end{figure}

\begin{figure}
\centering
\includegraphics[width=0.45\textwidth]{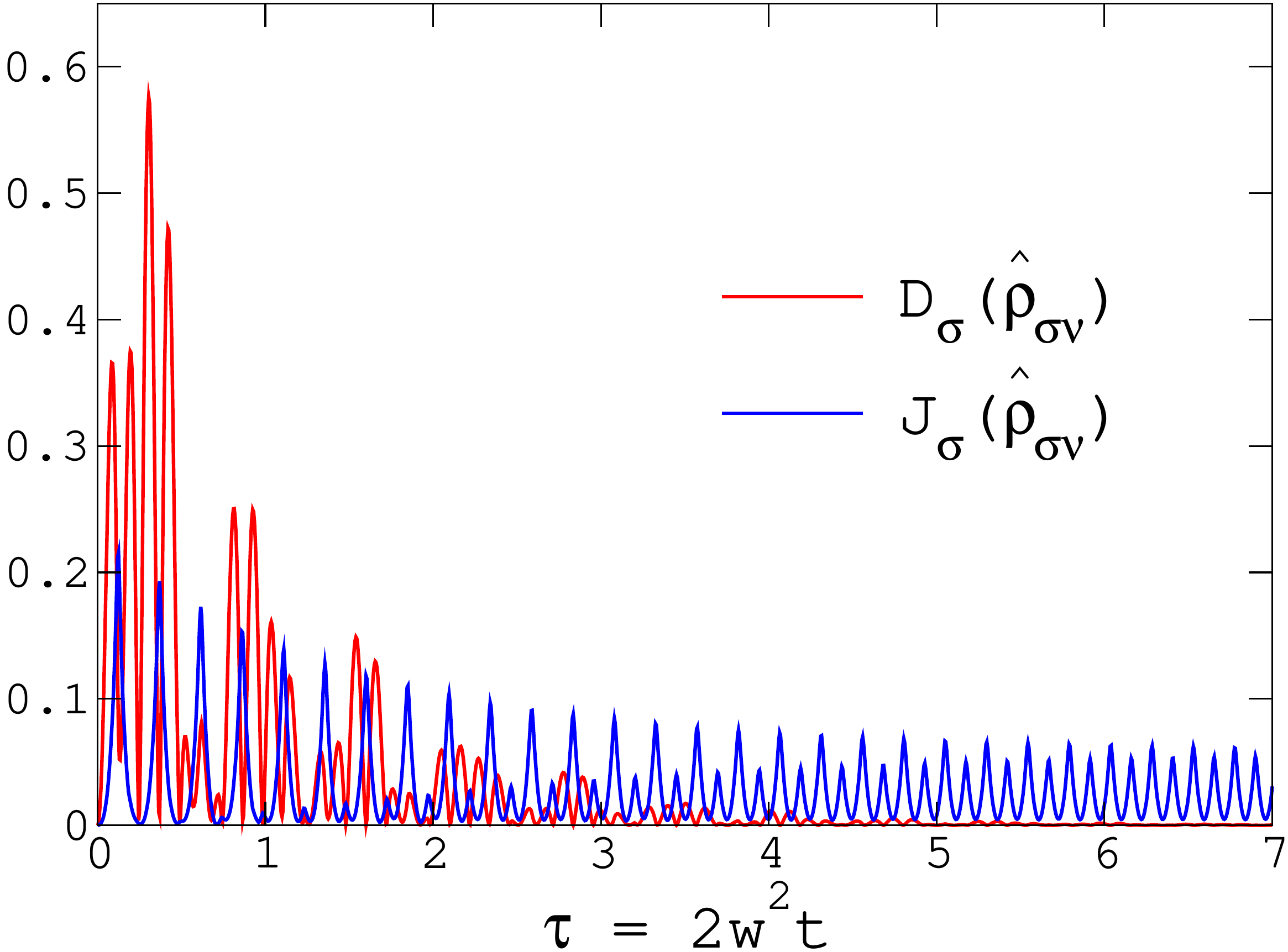}
\caption{
\label{fig:Discords1}
Classical correlation $J_\sigma (\hat \varrho_{\sigma \nu})$
corresponding to the maximal value
between $I(\sigma_z, \nu)$ and $I(\sigma_x, \nu)$ 
and quantum discord $D_\sigma (\hat \varrho_{\sigma \nu})$ expressed by
Eq. (\ref{quantum discord}),
at different values of $\tau$.
The time dependence of mutual informations is given in 
Figure \ref{fig:Informations1}.
}
\end{figure}

The mutual informations calculated with these entropies are shown in
Fig. \ref{fig:Informations1}.
According to Eq. (\ref{The classical correlation term}), the classical correlation is equal  to the maximum from the two  mutual informations $I(\sigma_z, \nu)$ and $I(\sigma_x, \nu)$.
As it can be seen, at early stage of evolution  $I(\sigma_x, \nu)>I(\sigma_z)$ and do not coincide with $I(\sigma, \nu)$. This fact confirms that quantum discord is not zero.
The classical correlation and discord are plotted
in Fig. \ref{fig:Discords1}. 
While the classical correlations show thermalized 
steady-state persistent oscillations, quantum discord shows a more non-trivial
behaviour.  
In the initial state, quantum discord is zero. 
It becomes non-zero in the transition stage of evolution and tends to zero 
again after thermalization. 

\begin{figure}[h!]
\centering
\includegraphics[width=0.45\textwidth]{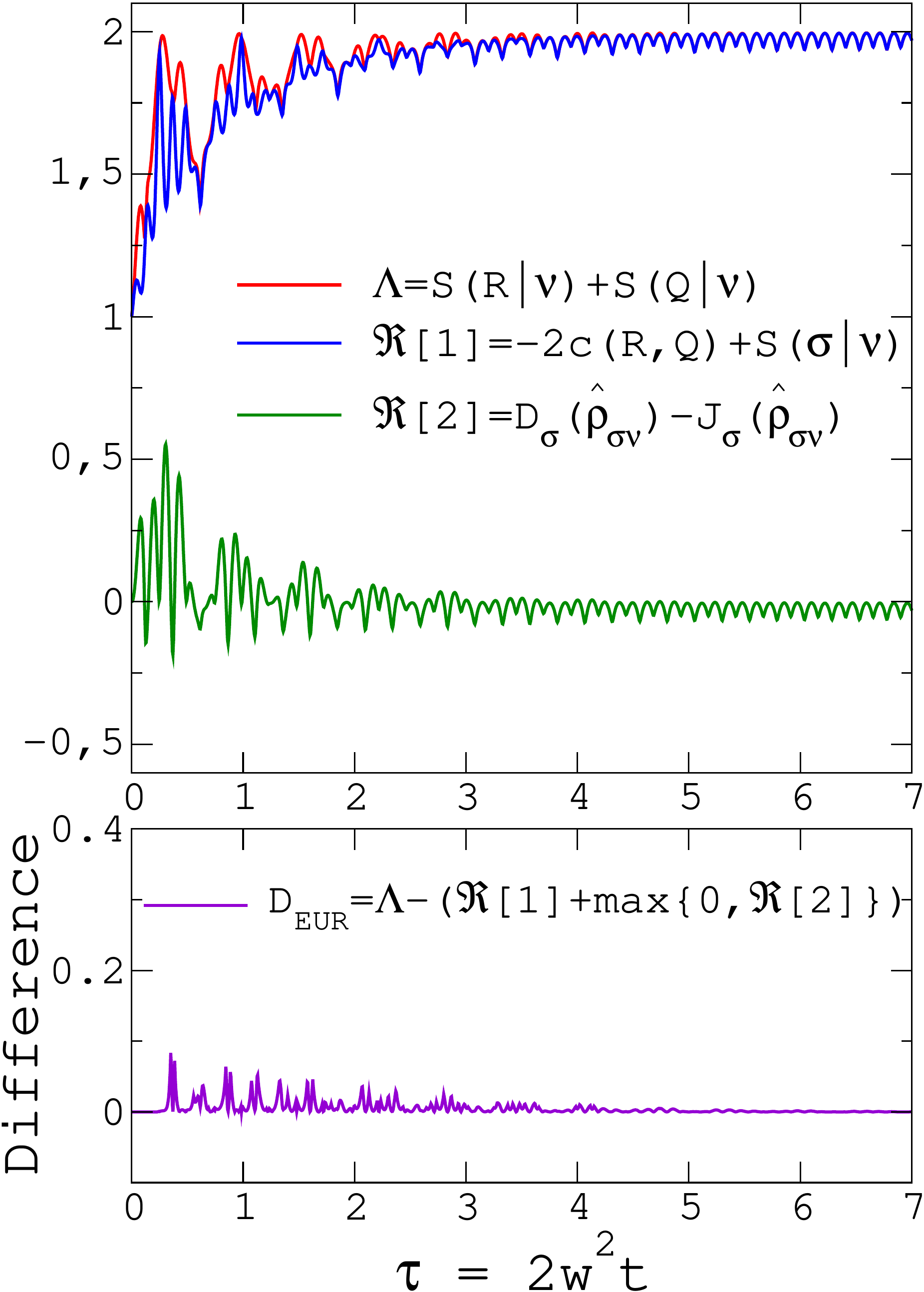}
\caption{
\label{fig:EUR1_difference}
Comparison of different terms in the uncertainty relation
(\ref{correction to the The Kraus}) (top), and
the difference between the left- and the right-hand sides (bottom)
at different values of the time parameter $\tau$.
$\Lambda$ corresponds to the left-hand side of the uncertainty relation,
while
$\mathcal{R}[1]$ is the sum of the right hand-side of
Eq. (\ref{trade-off relation1}) and the relative entropy $S(\sigma \vert \nu)$,
and $\mathcal{R}[2]$ is the difference between the quantum discord and
classical entropy.
As it is seen, there is a strict inequality when the system is not thermalized.
}
\end{figure}

After exploring the particular terms, we analyse the generalized Kraus's 
trade-off uncertainty relation Eq. (\ref{correction to the The Kraus}). 
The upper plot in Fig. \ref{fig:EUR1_difference} shows the comparison between 
entropies and the difference between the discord and classical correlation.  
The difference between the left- and right-hand sides of Kraus's trade-off uncertainty 
relation (\ref{correction to the The Kraus})   is shown in the bottom plot. 
The initial state corresponds to the zero discord. However, discord is generated very promptly, and after propagation on a long distance, discord again decays to zero.  In Fig. \ref{fig:EUR1_difference}, we see that in Eq. (\ref{correction to the The Kraus}), the inequality holds when discord is non-zero, and Eq. (\ref{correction to the The Kraus}) turns into equality after discord decays to zero. Thus we conclude that the discord reduces the lower bound of uncertainty, meaning that condition $D_{EUR}>0$ holds until discord is non-zero and the maximal value of discord corresponds to the maximum in $D_{EUR}$. After the discord decays to zero, inequality turns into equality $D_{EUR}=0$, meaning that uncertainty reaches its maximum possible value.

\section{\label{sec:conclusions}Summary and conclusions}

In the present work, we studied ultrahigh-energy neutrino propagation 
in the interstellar dissipative space.  
The mathematical problem reduces to the Lindbladian equation 
written for the density matrix of ultrahigh-energy neutrino. 
We considered two flavours of neutrino and the effect of spin-flavour 
oscillations. 
We obtained the exact analytical solution of the Lindbladian equation and 
analysed the effect of incompatible invasive measurements and 
uncertainty relations in particular. 
Incompatible measurements and corresponding uncertainty relations 
are still under incisive debate, even for non-relativistic systems.  
It is known that the best quantifiers of uncertainty are entropic 
measures which are defined differently and have various interesting features. 
The concept of a quantum memory was introduced recently and
modified in terms of trade-off relation. 
The trade-off relation contains many terms, such as marginal entropies
of the subsystems, quantum, and classical correlation terms. 
The quantum discord of the ultrahigh-energy neutrino plays the 
central role in our problem. While marginal entropies show similar behaviour, 
quantum discord defined through the difference between mutual information 
and classical correlations shows remarkable features.  
In the large time interval during the dissipative evolution, quantum discord is non-zero 
and reduces the uncertainty of incompatible 
measurements done on the ultrahigh-energy neutrino. 
This reduction is identified as an effect of the quantum spin-flavour memory 
hosted by the ultrahigh-energy neutrino.

\section*{\label{sec:datavail}Data Availability Statement}

Data sharing is not applicable to this article as no new data were created 
or analysed in this study.


%
\bibliographystyle{unsrt}
\bibliography{nems}
%
%

\end{document}